\documentclass[12pt]{article}

\usepackage{ltexpprt}

\usepackage[algo2e, ruled,vlined, linesnumbered]{algorithm2e}
\usepackage{multirow}
\usepackage{booktabs} 
\usepackage{subfig}
\usepackage{adjustbox}
\usepackage{dblfloatfix}
\usepackage{threeparttable}

\usepackage{cooltooltips} 
\usepackage{multido} 
\usepackage[colorinlistoftodos]{todonotes}

\usepackage{cite}
\usepackage{amsmath,amssymb,amsfonts}
\usepackage{algorithmicx}
\usepackage{graphicx}
\usepackage{textcomp}
\usepackage{xcolor}

\usepackage{pdflscape}

\def\BibTeX{{\rm B\kern-.05em{\sc i\kern-.025em b}\kern-.08em
		T\kern-.1667em\lower.7ex\hbox{E}\kern-.125emX}}

\definecolor{ForestGreen}{RGB}{63,142,38}
\definecolor{UnibasMint}{RGB}{30,165,165}

\definecolor{findNeighbors}{HTML}{0000FF}
\definecolor{Density}{HTML}{000075}
\definecolor{IAD}{HTML}{3791FF}
\definecolor{Divv}{HTML}{0B60FF}
\definecolor{Momentum}{HTML}{69CBFF}
\definecolor{Gravity}{HTML}{00AA00}
\definecolor{Communication}{HTML}{FF0000}
\definecolor{openMP}{HTML}{FE00FF}

\usepackage[normalem]{ulem}
\newcommand{\ac}[1]{{\color{black}#1}}

\newcommand{\ali}[1]{{\color{black}#1}}

\newcommand{\aliA}[1]{{\color{black}#1}}
\newcommand{\aliB}[1]{{\color{black}#1}}
\newcommand{\alir}[1]{{\color{black}#1}}

\newcommand{\dlbTool}{\textit{DLS4LB}}

\newcommand{\sphynx}{\mbox{SPHYNX}}
\newcommand{\tl}{thread level}
\newcommand{\pl}{process level}

\newcommand{\elap}{\textit{eLaPeSD}}

\newcommand{\cut}[1]{}
\usepackage{authblk}
\usepackage{lscape}
\begin{document}
\sloppy

\title{Two-level Dynamic Load Balancing for High Performance Scientific Applications}

\author[1]{Ali Mohammed}
\author[1]{Aur{\'e}lien Cavelan}
\author[1]{Florina M. Ciorba}	
\author[2]{	Rub{\'e}n M. Cabez{\'o}n}
\author[3]{Ioana Banicesu}

\affil[1]{Department of Mathematics and Computer Science\\
	University of Basel, Switzerland\\}
\affil[2]{	Scientific Computing Center (sciCORE), Department of Physics\\ University of Basel, Switzerland}
\affil[3]{	Department of Computer Science and Engineering\\ Mississippi State University, USA}

\renewcommand\Authands{ and }


 \maketitle
\clearpage

\tableofcontents
\clearpage

\begin{abstract}
\label{sec:abs}
Scientific applications are often complex, irregular, and \mbox{computationally-intensive}. 
To accommodate the \mbox{ever-increasing} computational demands of scientific applications, high performance computing~(HPC) systems have become larger and more complex, offering parallelism at multiple levels (e.g., nodes, cores per node, threads per core). 
Scientific applications need to exploit \emph{all} the available multilevel hardware parallelism to harness the available computational power.
The performance of applications executing on such HPC systems may adversely be affected by load imbalance at multiple levels, caused by problem, algorithmic, and systemic characteristics. 
Nevertheless, most existing load balancing methods do not simultaneously address load imbalance at multiple levels. 
This work investigates the impact of load imbalance on the performance of three scientific applications at the thread and process levels.
\emph{We jointly apply and evaluate selected dynamic loop \mbox{self-scheduling} (DLS) techniques to both levels}.
Specifically, we \ali{employ} the extended LaPeSD OpenMP runtime library~\cite{Ciorba:2018} at the \tl{}, and \ali{extend} the \dlbTool{} \mbox{MPI-based} dynamic load balancing library~\cite{companion_research_report} at the \pl{}.
This approach is generic and applicable to any \mbox{multiprocess-multithreaded} \mbox{computationally-intensive} application (programmed using MPI and OpenMP).
We conduct an exhaustive set of experiments to assess and compare six DLS techniques at the \tl{} and eleven at the \pl{}. 
The results show that improved application performance, by up to $21\%$, can only be achieved by \emph{jointly} addressing load imbalance at the two levels.
We offer \ali{insights into} the performance of the selected DLS techniques and discuss the interplay of load balancing at the \tl{} and \pl{}. 

\end{abstract}


\textbf{Keywords.}\\
Two-level dynamic load balancing, \mbox{Computationally-intensive} applications, High performance computing, \mbox{Self-scheduling}, MPI+OpenMP

\clearpage
\section{Introduction}
\label{sec:intro}



\aliA{Scientific applications are typically large, irregular, and \mbox{computationally-intensive}.
To match the \mbox{ever-increasing} computational demands of scientific applications, high performance computing (HPC) systems have become larger and more complex. 
Not only the number of compute nodes in an HPC system has increased, but also the number of CPU sockets and the number of cores per socket have increased. 
For example, the number of CPU cores per node in the top 3 supercomputing systems\footnote{https://www.top500.org/lists/2018/11/} is ranging from $44$ to $260$ cores.
Therefore, applications need to exploit hardware parallelism at multiple levels to achieve the best performance.

Due to the hybrid nature of current HPC systems, distributed memory across compute nodes and shared memory within a single node, hybrid parallelization of applications at \pl{} and \tl{} using \mbox{MPI+OpenMP} is the most common and successful approach in scientific applications~\cite{jin2011high, rabenseifner2009hybrid, smith2001development}.  
However, the performance of scientific applications on such systems may be degraded due to load imbalance.
Load imbalance can be caused by irregular application or computing system characteristics, such as conditional statements leading to the variation of computations and \mbox{non-uniform} memory access latency, respectively.
Load imbalance degrades application performance and hinders its scalability.
}
\aliA{To balance the load among parallel processes, several scientific applications use domain decomposition methods, such as octagonal recursive bisection (ORB)~\cite{kandhai1999large}.
	However, the computational load per sub-domain may change as the execution evolves and causes imbalance. 
	Also, these methods can not adapt to the load imbalance due to system characteristics, such as \mbox{non-uniform} memory \ali{accesses} (NUMA) and perturbations.
	Other applications, such as ChaNGa~\cite{menon2015adaptive}, rely on load balancing solutions offered by specific programming abstractions (Charm++)\footnote{http://charm.cs.illinois.edu/research/charm}.
	In addition, applications, such as Lassen, Kripke, and ChaNGa, create OpenMP tasks to help overloaded chares (in Charm++) to balance the load among processing elements (PEs) in a shared memory domain~\cite{bak2018multi}. 
	Also, as the number of PEs in a shared memory domain has increased, the standard OpenMP scheduling options may not be sufficient to achieve a balanced load execution~\cite{Ciorba:2018}.
}

\begin{figure}[htbp]
	\centering
	\includegraphics[clip, trim=0cm 0cm 0cm 0cm, scale=0.9]{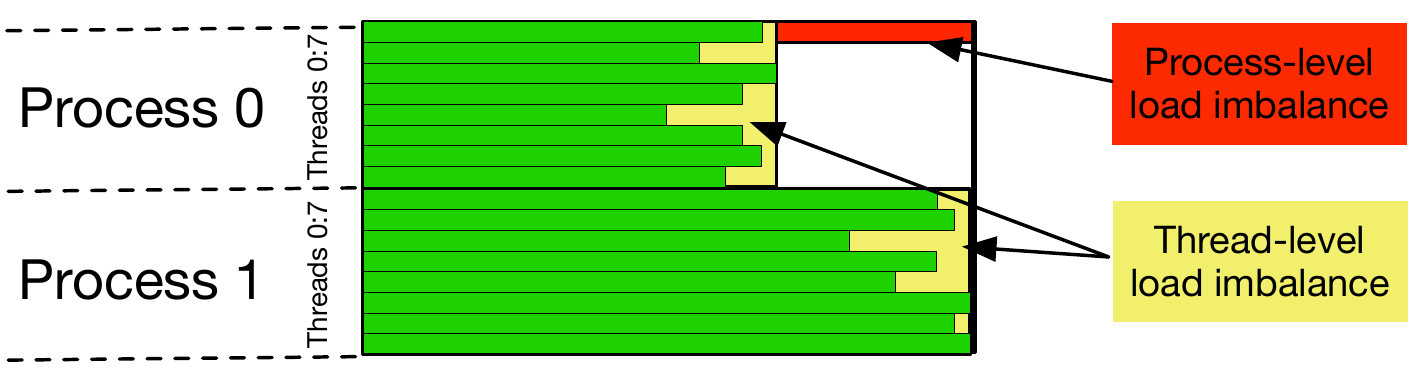}%
	\caption{Conceptual illustration of the impact of the \mbox{two-level} load imbalance of a scientific application on two processes each with eight threads. Due to the uneven load at the \tl{}, \aliA{faster} threads wait \aliA{for} the slowest thread in each process, represented by the yellow regions. At the \pl{}, process~0 \aliA{(faster)} waits for process~1 \aliA{(slower)}, represented \aliA{by} the red region. The application completes when the slower process (process 1) finishes.}
	\label{fig:two-level-hl}
\end{figure}

Moreover, load imbalance may manifest in more than one level of software parallelism, i.e., among processes (\mbox{process-level}) and among threads (\mbox{thread-level}).
For example, \figurename{~\ref{fig:two-level-hl}} conceptually shows the \mbox{two-level} load imbalance of a scientific application parallelized using multiple processes and threads.
Due to the \tl{} load imbalance, threads that finish early \aliA{must} wait until the slowest thread finishes (yellow regions).
Therefore, the performance of a process is dominated by \aliA{its} slowest thread. 
Similarly, at the \pl{}, the faster process has to wait for the slower one and the application performance is dominated by \aliA{its} slowest process.
The \mbox{two-level} load imbalance is a compound problem and not trivial to address as the scheduling \aliA{performance} at one level is \aliA{influenced} by the scheduling decisions \aliA{at} the other. 
For example, the relation between batch and application level scheduling was studied~\cite{Eleliemy:2017a}, and it was shown that a holistic solution results in better \aliA{performance improvement} than focusing on improving the performance at each level alone.

In this work, dynamic load balancing via dynamic loop \mbox{self-scheduling} (DLS) is jointly applied at both the \mbox{thread-} and the \mbox{process-levels} to achieve improved performance.
DLS \mbox{self-schedules} groups of loop iterations \aliA{(tasks)}, or chunks, to free and requesting PEs, e.g., processes \aliA{or} threads,  to achieve \ali{a \aliA{dynamically} balanced \aliA{load} execution}.
DLS techniques have \ali{successfully} been used to achieve load balance in \mbox{computationally-intensive} scientific applications, such as heat diffusion~\cite{banicescu2001load}, wave packet simulations~\cite{AWFBC}, and \mbox{N-Body} simulations~\cite{fractiling}.
\aliA{The} DLS techniques are generic \ali{and} \aliA{have been} implemented in various programming models.
In this work, they are applied at the thread- and the \mbox{process-levels} to three scientific applications, Mandelbrot~\cite{mandelbrot1980fractal}, PSIA~\cite{psia}, and \sphynx{}~\cite{cabezon2017sphynx}.
DLS techniques are applied to the three applications at the \tl{} using an extended version of \ali{the} GNU OpenMP runtime time library~\cite{Ciorba:2018} (\elap{}), and at the \pl{} using an extended version of  the \dlbTool{}~\cite{companion_research_report}. 
Sixty six combinations of DLS techniques at the thread and the process levels are \ali{explored} to \ali{find} the best combination of DLS techniques at both levels for the applications of interest.

This work makes the following contributions: 
(1)~\ali{A} generic approach of \mbox{two-level} \aliA{dynamic} load balancing using DLS techniques with \elap{} at the \tl{}, and the extended \dlbTool{} at the \pl{};
(2)~The extension of the \dlbTool{}~\cite{companion_research_report} with \ali{the} adaptive weighted factoring (AWF) technique that supports \mbox{time-stepping} applications; the modification of the \ali{AWF} variants namely AWF-B, C, D, E to share the learned \aliA{updated} weights between \mbox{time-steps}; 
(3)~The analysis of the \mbox{two-level} load imbalance in \aliA{three scientific applications} and its impact on performance;
(4)~Exhaustive scheduling experiments \ali{that} test different combinations of \tl{} and \pl{} scheduling techniques that led to the performance enhancement of applications up to $21\%$; 
(5)~Insights into the performance of the selected DLS techniques and the interplay of load balancing at the \tl{} and \pl{}.

This work is structured as follows: Section~\ref{sec:background} contains a brief review of \mbox{self-scheduling} techniques, as well as of the work related to the load balancing in \ali{the} literature and the use of DLS techniques to balance the load of scientific applications. 
The proposed methodology of using DLS techniques to load balance at the two (thread and process) levels is explained in Section~\ref{sec:dlb_in_SPHYNX}. 
The experimental design and setup and the performance of the proposed approach are described and discussed in Section~\ref{sec:results}. 
The work concludes and outlines potential future work in Section~\ref{sec:conc}.


\section{Background and Related Work}
\label{sec:background}
%
A brief background on \aliA{\mbox{self-scheduling}} techniques is presented in this section. 
Besides, load balancing libraries and the applied load balancing methods to \aliA{scientific applications} are discussed.

\subsection{Background}
\label{subsec:background}
\subsubsection{Self-scheduling}
\aliA{The iterations of \mbox{computationally-intensive} loops or tasks are assigned} to PEs to achieve a balanced load execution with minimum overhead.
Loop scheduling techniques are divided into static and dynamic.
In this work, block scheduling is considered, \ali{denoted as STATIC,} where each PE is assigned a block \aliA{or} chunk of tasks equal to the number of tasks, $N$, divided by the number of PEs, $P$. 
DLS techniques assign a chunk of tasks to free and requesting PEs during execution via \emph{\mbox{self-scheduling}}.
\aliA{\mbox{Self-scheduling} is different from \emph{work~stealing}~\cite{blumofe1999scheduling} where PEs are initially assigned a block of tasks and they need to steal afterward form distributed work queues to balance the load, whereas in \mbox{self-scheduling} tasks are only assigned to free and requesting PEs from a central work queue.}
\aliA{The} DLS techniques can further \aliA{be} divided into nonadaptive and adaptive techniques. 
\aliA{The} nonadaptive DLS techniques address the load imbalance \ali{caused by} problem or application characteristics, such as the variation of \aliA{tasks} execution times. 
Nonadaptive DLS techniques include self scheduling~\cite{SS}~(SS), \mbox{fixed-sized} chunking~\cite{FSC}~(FSC), modified \mbox{fixed-sized} chunking~\cite{banicescu:2013:a}~(mFSC), guided \mbox{self-scheduling}~\cite{GSS}~(GSS), trapezoid \mbox{self-scheduling}~\cite{tzen1993trapezoid}~(TSS), factoring~\cite{FAC}~(FAC), weighted factoring~\cite{WF}~(WF), and random~\cite{Ciorba:2018}~(RAND).
SS assigns a single loop iteration at a time per PE request. 
Thus, it results in the maximum load balance and the maximum scheduling overhead.
SS represents one extreme, where the load balancing \ali{effect} and the scheduling overhead \ali{are at} maximum, \ali{whereas} STATIC represents the other extreme, \ali{where load balancing effect and the scheduling overhead are at minimum}.
FSC assigns loop iterations in chunks of fixed size, \ali{hence reducing} the scheduling overhead compared to SS. 
The chunk size depends on the scheduling overhead, $h$, and the standard deviation of the iterations execution time, $\sigma$.
mFSC alleviates the burden of determining $h$ and $\sigma$ and assigns a chunk size that results in a number of chunks that is similar to that of FAC (explained below).
GSS addresses the uneven starting times \ali{of PEs} and assigns chunks in decreasing sizes. 
The chunk \ali{sizes in GSS are calculated as} the number of the remaining loop iterations, $R$, divided by $P$.
TSS assigns chunks of decreasing sizes, similar to GSS. 
However, chunk sizes decrease linearly in TSS, \ali{which simplifies} the chunk calculation and reduces scheduling overhead.
FAC assigns chunks in batches to reduce the scheduling overhead.
FAC \ali{employs} probabilistic analysis of application characteristics to calculate batch sizes that maximize the probability of achieving a balanced load execution.
The batch size calculation depends on the mean of iterations execution times, $\mu$, and their standard deviation, $\sigma$. 
The chunk sizes are equal \ali{in} a batch, \ali{namely} the batch size divided by $P$.
When $\mu$ and $\sigma$ are not available, FAC is practically implemented by assigning half of the remaining loop iterations \ali{as} a batch, \ali{which is equally distributed to PEs on request}.
WF is similar to FAC, except that it addresses heterogeneous PEs.
In WF, each PE is assigned a relative weight that is fixed during execution. 
Each PE is assigned a chunk \ali{from the current} batch relative to its weight.
In this work, the practical implementations of FAC and WF are used.
RAND employs the uniform distribution to arrive at a randomly calculated chunk size between an upper and a lower bound. 
The randomly calculated chunk size is bounded by $ N/ (100 \times P) \le chunk\_size \le N/(2 \times P)$~\cite{Ciorba:2018}.
\aliA{The} adaptive DLS techniques measure the performance during execution and adapt \ali{their} chunk calculation accordingly to address the load imbalance due to systemic characteristics, such as \mbox{non-uniform} memory access (NUMA) delays and perturbations during execution.  
\aliA{The} adaptive DLS techniques include adaptive \ali{weighted} factoring~\cite{AWF}~(AWF), its variants~\cite{AWFBC}~AWF-B, AWF-C, AWF-D, AWF-E, and adaptive factoring~\cite{AF}~(AF), among others.
AWF adapts the relative PE weights during execution according to their performance.
It is designed for \mbox{time-stepping} applications. 
It measures the performance of PEs during previous \mbox{time-steps} and updates the PEs relative weights after each \mbox{time-step} to balance the load according to the computing system's \ali{present} state.
\mbox{AWF-B} relieves the \mbox{time-stepping} \ali{requirement} to learn the PE weights.
It learns the PE weights from their performance in previous batches instead of \mbox{time-steps}.
\mbox{AWF-C} is similar to \mbox{AWF-B}, however, the PE weights are updated after the execution of each chunk, instead of batch.
\mbox{AWF-D} is similar to \mbox{AWF-B}, where the scheduling overhead (time taken to assign a chunk of loop iterations) is taken into account in the weight calculation. 
\mbox{AWF-E} is similar to \mbox{AWF-C}, and takes into account also the scheduling overhead, similar to \mbox{AWF-D}.
AF is \ali{also} based on FAC.
\ali{However,} it measures the performance of PEs to learn \ali{the} $\mu$ and $\sigma$ per PE during execution.

\subsubsection{DLS implementation in MPI and OpenMP}
A dynamic load balancing tool (\mbox{\textit{DLB\_tool}}) \aliA{that} implements certain DLS techniques at the \pl{} using MPI was introduced and used to balance the load of an image denoising model and at \ali{the} simulation of a vector functional coefficient autoregressive (VFCAR) model for multivariate nonlinear time series~\cite{carino2007tool}.
The \mbox{\textit{DLB\_tool}} initially implemented nine loop scheduling techniques: STATIC, mFSC, GSS, FAC, AWF-B, AWF-C, AWF-D, AWF-E, and AF. 
The \aliA{\mbox{\textit{DLB\_tool}}} \ali{employ} \mbox{self-scheduling} and use a \mbox{master-worker} execution model. 
Free MPI processes request work from the master \aliA{process} that \ali{executes} the \mbox{self-scheduling} techniques. 
The master \ali{also} doubles as a worker \aliA{process} and executes chunks of tasks.
The \mbox{\textit{DLB\_tool}} was further extended into \ali{\dlbTool{}}~\cite{companion_research_report} with four \aliA{additional} DLS techniques: SS, FSC, TSS, and WF \aliA{to support $13$ DLS techniques in total,} and was used to balance the load of two scientific applications (PSIA~\cite{Eleliemy:2017b} and Mandelbrot~\cite{mandelbrot1980fractal}) and five synthetic workloads.

\ali{At} the \tl{} scheduling, the GNU OpenMP runtime library was extended \aliA{into}~\elap{}~\cite{Ciorba:2018} to support \ali{four} additional DLS techniques: FSC, TSS, FAC, and RAND \aliA{in addition to} the \aliA{standard} OpenMP scheduling techniques: STATIC, dynamic~(\aliA{actually} SS~\cite{SS}), and guided~(\aliA{actually} GSS~\cite{GSS}).

\subsubsection{Single-level dynamic load balancing}
DLS techniques have been used in several studies to improve the performance of \mbox{computationally-intensive} scientific applications \aliA{at a single software parallelism level: \pl{}}. 
For example, \aliA{SS, FAC, AWF, and AF} \ali{were} used to balance the load of \ali{a} heat conduction application on an unstructured grid~\cite{banicescu2001load}.  
AF was found to result in a superior performance, especially with irregular applications \ali{executing} in heterogeneous environments.
\ali{The} DLS techniques \ali{were} used to balance the load of scientific applications, such as simulations of wave packet dynamics,  N-Body simulations~\cite{fractiling}, automatic quadrature routines~\cite{AWFBC}, and a computer vision application (PSIA)~\cite{Eleliemy:2017b}.
\aliA{The DLS techniques were also used to balance the load of scientific benchmarks, such as NAS parallel~\cite{bailey2011parallel} and RODINIA~\cite{che2009rodinia} benchmarks, at the \tl{}~\cite{Ciorba:2018}.}  
\aliA{This work is the first to jointly explore the use of the DLS techniques at both process- and thread-levels.}

\subsection{Related Work}
\label{subsec:related_work}
\aliA{A hierarchical domain decomposition using \mbox{space-filling} curves was introduced to achieve a balanced load execution of an atmospheric cloud model~\cite{lieber2018highly}.
The division of the \mbox{space-filling} curve and the assignment of each part are performed in two stages using an exact algorithm and a heuristic to reduce the complexity of the curve cutting problem to achieve load balancing with low overhead.
However, the above work used hierarchical scheduling to simplify the scheduling problem at a single software parallelism level.
Hierarchical scheduling has been used also at the \tl{} for the scheduling on \mbox{multicore} \mbox{hyper-threaded} CPUS, where the first level of the hierarchical scheduling is for scheduling work to the cores, and the second level is for scheduling and load balancing between hyper threads that share the same core.
However, no interaction between the scheduling on the two levels were considered or studied and only studied hierarchical scheduling within one level of parallelism, i.e. \tl{}.}
Quo~\cite{gutierrez2018adaptive} was introduced to adapt threads and processes binding during runtime to improve MPI+OpenMP applications performance.
\aliA{Quo adapts processes/threads bindings to PEs to accommodate the newly spawned/suspended threads during different computational phases within an application and preserve data locality.
However, an overloaded thread or process may still cause load imbalance at the \tl{} or the \pl{}.}
ChaNGa~\cite{menon2015adaptive} uses over-decomposition, supported by Charm++, to allow fine and dynamic load balancing of chares execution among PEs.
Load statistics are collected, and \ali{particles} are migrated between the PEs to balance the load~\cite{menon2012automated}.
OpenMP was integrated with Charm++ to balance the load among PEs in a shared memory domain in applications, such as Lassen, Kripke, and ChaNGa~\cite{bak2018multi} \aliA{by creating OpenMP tasks to help overloaded chares}. 
Adaptive hierarchical scheduling~\cite{wang2014adaptive}~(AHS) was proposed for the \mbox{two-level} scheduling on multi-core clusters. 
The work is initially divided between the nodes with the aim to achieve a balanced load among them.
\mbox{Work-stealing} was used for \mbox{intra-node} load balancing between the threads whereas both \mbox{work-stealing} and \mbox{work-sharing} were used for \mbox{inter-node} load balancing.

Load balancing solutions based on domain decomposition can not adapt to accommodate \aliA{all (unpredictable)} variations in the computing system characteristics.
Balancing solutions that are \mbox{language-specific}, such as Charm++, can not easily be ported to other \aliA{scientific applications}.
None of the \aliA{efforts} mentioned above analyzed the \aliA{effects of} two-level \aliA{(process and thread)} load imbalance \ali{nor} studied the \aliA{benefits} of load balancing at one \ali{of the levels} on the other level.

\ali{This} work \ali{studies} load imbalance at the \tl{} and the \pl{} in \aliA{3 scientific applications; PSIA~\cite{psia}, Mandelbrot~\cite{mandelbrot1980fractal}, and \sphynx{}~\cite{cabezon2017sphynx}.}
The \elap{} is used to balance the load at the \tl{} and the \dlbTool{}~\cite{companion_research_report} to support \aliA{the} AWF DLS technique (that supports \mbox{time-stepping} applications, such as \sphynx{} in this work), to balance the load at the \pl{}.
Moreover, adaptive \aliA{DLS} techniques, such as \aliA{the} AWF-B, AWF-C, AWF-D, and AWF-E are \aliA{improved} such that the learned \aliA{updated} \aliA{PE} weights are transferred from previous \mbox{time-steps} to \ali{the} current \mbox{time-step}, to support \mbox{time-stepping} applications, such as \sphynx{}. 
\aliA{Sixty six joint} combinations \ali{of} DLS techniques at the \tl{} and the \pl{} are \ali{tested} to achieve the best performance for the three scientific applications. 

\section{Two-level Dynamic Load Balancing} 
\label{sec:dlb_in_SPHYNX}

\begin{figure*}[htbp]
	\centering
	\includegraphics[clip, trim=0cm 0cm 0cm 0cm, scale=0.6]{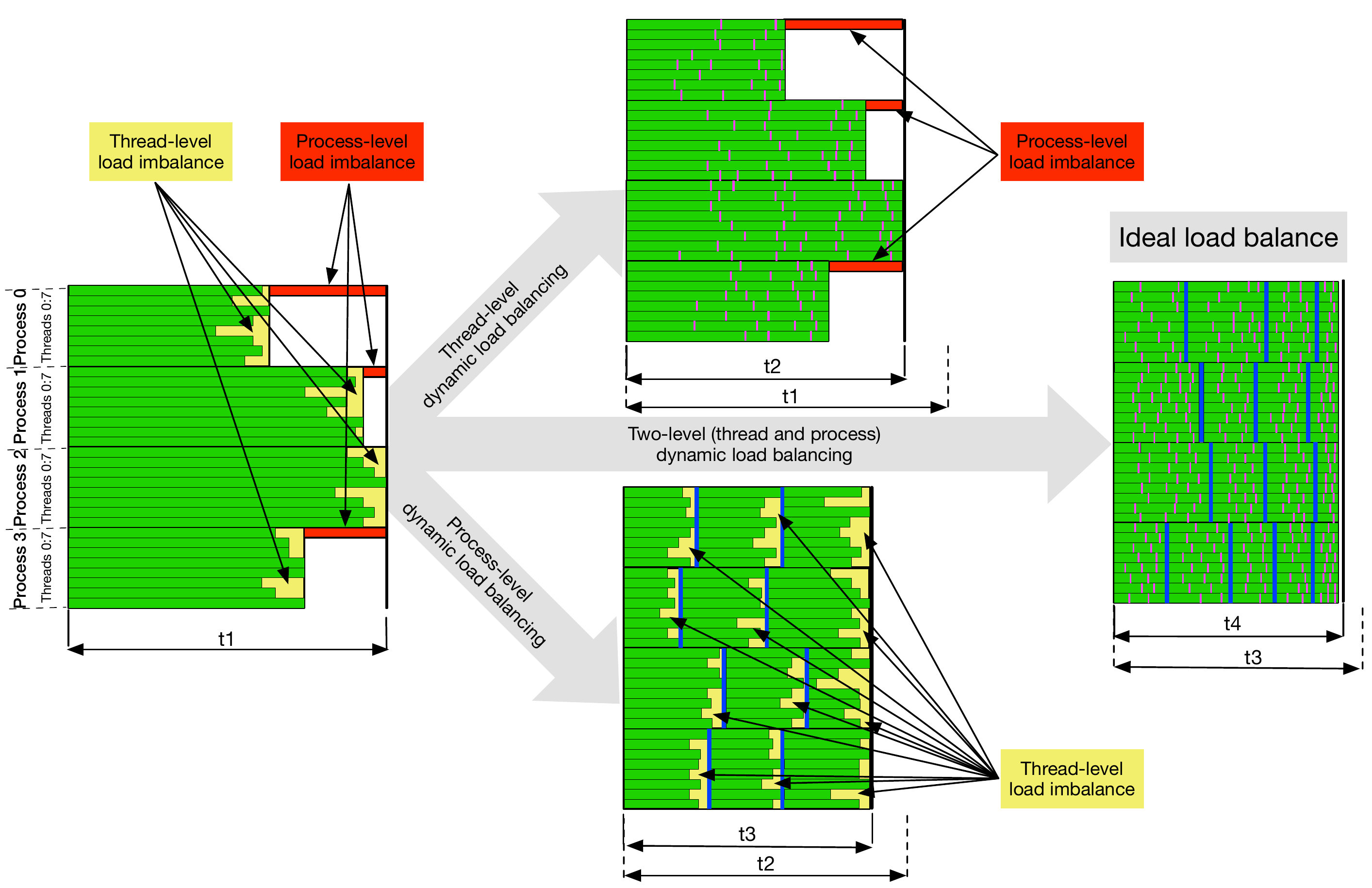}%
	\caption{Conceptual illustration of \aliA{employing} two-level \aliA{dynamic} load balancing via \elap{} and  \dlbTool{}. At the \tl{}, OpenMP threads are \aliA{\mbox{self-scheduled}} chunks of \aliA{tasks} using the extended GNU runtime library \elap{}. At the \pl{}, the \dlbTool{} \aliA{\mbox{self-schedules}} chunks of \aliA{tasks} to free and requesting MPI ranks \ali{in multiple rounds}.}
	\label{fig:two-level}
\end{figure*}

Addressing load imbalance at both the process and the thread levels is essential to achieve improved application performance.
\figurename{~\ref{fig:two-level} describes the \ali{proposed} two-level \aliA{dynamic} load balancing \ali{via \mbox{self-scheduled} approach}. 
At the \pl{}, the \dlbTool{} \aliA{\mbox{self-schedules}} a chunk of \aliA{tasks} to a free and requesting MPI rank \ali{in multiple rounds}.
The work assigned to an MPI rank is parallelized and distributed among \aliA{several} OpenMP threads (8 threads per rank) using the \elap{} OpenMP runtime library \ali{for load balancing}.
\aliA{The library \mbox{self-schedules}} a chunk of \aliA{tasks} (\aliA{a subset of the loop iterations assigned to} this MPI rank \ali{at} the \pl{} by the \dlbTool{}) to \ali{a} free and requesting thread.
\aliA{Employing DLS at only one level (either \pl{} or \tl{}) achieves load balance at this level and improves the application performance as shown in the middle subplots in~\figurename{~\ref{fig:two-level}}.
However, the best application performance can only be achieved by employing dynamic load balancing at the two levels as shown in~\figurename{~\ref{fig:two-level}}.}
Based on the selected DLS technique at the \tl{} \ali{via} \aliA{the} \texttt{OMP\_SCHEDULE} \aliA{environment variable,} and the selected scheduling technique at the \pl{}, \ali{via} the \dlbTool{}, along with the application and the computing system properties, different degrees of \ali{load balancing} and performance \aliA{improvement} are achieved as shown in Section~\ref{sec:results}.
The two-level \aliA{dynamic} load balancing approach \ali{proposed and} depicted in \figurename{~\ref{fig:two-level}} is generic and can be used with any scientific application parallelized with MPI+OpenMP hybrid parallelization.

\subsection{Implementation}
To balance the load at the \tl{}, an extended version of GNU OpenMP \ali{runtime} library LaPeSD, i.e., \elap{}~\cite{Ciorba:2018} is used. 
\elap{} supports seven OpenMP scheduling techniques that can be selected by exporting the name of the scheduling technique to the OpenMP environment variable \texttt{OMP\_SCHEDULE}.
\aliA{Recall that} \elap{} provides four additional dynamic loop scheduling techniques, FSC, TSS, FAC, and RAND \ali{in addition to} the three original OpenMP scheduling techniques: static, dynamic, and guided.
\aliA{The} OpenMP static scheduling \aliA{clause corresponds to the} STATIC \aliA{technique}, where each thread is assigned only one chunk of size $N$/$P$.
The dynamic and guided \aliA{scheduling clauses correspond to} SS and GSS, respectively~(c.f. Section~\ref{sec:background}).
To enable the application to read and use scheduling algorithms defined in the OpenMP runtime library, \texttt{schedule(runtime)} \aliA{needs to} be added to \ali{the} OpenMP parallelization of a for loop (in C) or a do loop (in FORTRAN), \ali{as shown in Listing~\ref{algo:two-level} Line~$21$~(line in magenta font color)}.
The path to the \elap{} \ali{library} \aliA{needs to} be added to the dynamic library path variable \texttt{LD\_LIBRARY\_PATH} to call \aliA{our} extended \ali{OpenMP runtime} library instead of the standard one.
Currently, \aliA{our} extended library only works with GNU compilers, \aliA{while} an extended LLVM OpenMP \ali{runtime} library is currently under development.

To balance the load at the \pl{}, \aliA{our extended} \dlbTool{} is used to \aliA{dynamically} distribute the \aliA{application tasks} to MPI ranks \ali{in multiple rounds} via \mbox{self-scheduling}. 
\aliA{Recall that} the \dlbTool{} \ali{provides} 13 loop scheduling techniques, ranging from \ali{fully} static to \ali{fully} dynamic, nonadaptive and adaptive, namely: STATIC, SS, FSC, mFSC, TSS, GSS, FAC, WF, AWF-B, AWF-C, AWF-D, AWF-E, and AF~(c.f. Section~\ref{sec:background}). 
In this work, it is extended to support an additional DLS technique; the AWF originally developed for \mbox{time-stepping} applications~\cite{AWF}.
This is needed due to the \mbox{time-stepping} nature of scientific simulations, such as \sphynx{}.
Moreover, AWF-B, AWF-C, AWF-D, and AWF-E are \aliA{improved} to share the learned \ali{updated} \aliA{PE} weights  between \mbox{time-steps}, instead of starting with a test chunk at each \mbox{time-step} to estimate the PE weights. 
\ali{Calls to the \dlbTool{} need to be inserted before and after the main calculations \aliA{such that they are \mbox{self-scheduled} using} DLS techniques \aliA{as shown by lines in blue font color in Listing~\ref{algo:two-level}}.
In addition, steps for setting up and finalizing the \dlbTool{}, such as data allocation, deallocation and selecting the DLS technique, need to be added before and after the \mbox{time-stepping} loop (Lines~$9-14$), respectively.}

The application needs to be a \mbox{time-stepping} application only to use the AWF technique. 
\aliA{To use} the \dlbTool{}, the application data \aliA{need to} be replicated among processes as the current implementation of the \dlbTool{} does not communicate the data required to compute the assigned chunk of loop iterations.
It is the programmer's responsibility to \ali{ensure} that the data are available for computation and \ali{are} communicated correctly.
\ali{To use} the extended version of the OpenMP runtime library, the application \aliA{needs to} be compiled with \aliA{the} GNU compilers, and the path to the extended library should be exported before execution.

\begin{algorithm2e}[htbp]
	\caption{Two-level Dynamic Load Balancing via \mbox{Self-scheduling}}
	\label{algo:two-level}
	\#include <mpi.h>\\
	\#include <omp.h>\\
	{\color{blue} \#include ``\dlbTool{}.h''}\\
	
	\texttt{int main()}\\
	\texttt{\{}\\
	/* Application initialization*/ \\
	\dots\\
	{\color{blue}  DLS4LB\_setup($P$, $N$, DLS\_method);}\\
	
	\For{$l\gets t_{init}$ \KwTo $t_{final}$}
	{
	
		\dots\\
		{\color{blue} DLS4LB\_Start\_loop();}\\
	 	\textbf{Main\_calculations();}\\
		{\color{blue} DLS4LB\_End\_loop();}\\
		\dots \\
	}
	{\color{blue} DLS4LB\_Finalize();}\\
	\texttt{\}} /* End main */\\
	
\vspace{0.5cm}

	\texttt{void Main\_calculations()}\\
	\texttt{\{}\\
	{\color{blue} \While{ {\color{blue} ! DLS4LB\_Terminated()}}
		{
			{\color{blue} DLS4LB\_Start\_chunk($loop_{start}$, $loop_{end}$);}\\
			{\color{black} 
				\#pragma omp parallel for {\color{openMP}schedule(runtime) }\\
				\For{$i \gets loop_{start}$ \KwTo $loop_{end}$ }
				{
					\textbf{/* Execute loop body */}\\
					\dots\\
				}
			} 
			{\color{blue} DLS4LB\_End\_chunk();}\\
		} 
	} 
	\texttt{\}} /* end main calculations*/\\
\end{algorithm2e}	

\subsection{Execution}
Each MPI rank (process) sends a work request to the master rank when it becomes free (Algorithm~1, Line~20) using the \dlbTool{}.
In response, the master rank assigns a chunk of tasks to the requesting MPI rank.
The size of the assigned chunk is determined by the employed DLS technique (specified in Algorithm~1, Line~8).
\emph{This allocated chunk at the MPI level is subsequently distributed to OpenMP threads for further scheduling and execution at the thread level.}
Therefore, threads are assigned chunks (or \mbox{sub-chunks} of the chunk allocated at the MPI level) of tasks whenever they become free using \elap{}.
Threads are assigned work until they complete the execution of the chunk allocated to their respective MPI rank.
The process repeats until all tasks ($N$~tasks) complete 
and \texttt{\mbox{Main\_calculations}} of the current \mbox{time-step} is completed.

An application needs to be a \mbox{time-stepping} application to use the AWF technique. 
Otherwise, an application may use all other DLS techniques available in the \dlbTool{} library.
The current implementation of the \dlbTool{} does not distribute application's data.
Applications need to ensure that the data are available where the work is assigned either by replicating the data or communicating the data with work.

Calls to the dynamic load balancing libraries, such as \elap{} or \dlbTool{}, incur overhead, compared to using a static scheduling approach.
This overhead is proportional to number of scheduling rounds and the cost of chunk calculation, which depends on the scheduling technique.
In \mbox{two-level} dynamic load balancing, this overhead is proportional to the product of the number of scheduling rounds at each of the two levels.
However, in severe load imbalance cases, this overhead is unavoidable and expected to be absorbed by the performance gain resulting from dynamic load balancing (c.f. Section~\ref{sec:results}).


%
\section{Performance Evaluation and Discussion} 
\label{sec:results}

 \begin{table}[!h]
		\caption{Details used in the design of factorial experiments \aliA{for performance analysis}.}
	\begin{center}
		\label{tbl:design_of_exp}
		\begin{adjustbox}{max width=\textwidth}
			\begin{threeparttable}
				\begin{tabular}{@{}lll@{}}
					\toprule
					\textbf{Factors}   & \textbf{Values}  & \textbf{Properties}  \\ \midrule
					
					\multirow{3}{*}{ \begin{tabular}[c]{@{}l@{}}  \\ \\ \textbf{Applications}  \end{tabular} }      & Mandelbrot \aliB{(Mathematics)} &  $N=0.6 \times 10^6$ tasks \\ \cline{2-3}
					&    PSIA \aliB{(Computer vision)}      &  $N=0.8\times 10^6$ tasks \\ \cline{2-3}
					&    \sphynx{} \aliB{(Astrophysics)}  &    \begin{tabular}[c]{@{}l@{}} Test-case (1): Stellar collision, $N= 10.4\times 10^6$ tasks\\ \hspace{2.1cm} Time-step: $6900$\\ Test-case (2a): Evrard collapse, $N = 1\times 10^6$ tasks\\ \hspace{2.1cm} Time-step: $100$, $500$, $1000$, $1700$, $2000$, $2300$, $2500$, $2800$, full simulation[$0:3000$]\\ \alir{Test-case (2b): Evrard collapse, $N = 1\times 10^7$ tasks}\\ \alir{\hspace{2.1cm} Time-step: $1$} \end{tabular} 
					\\ \midrule	      
					\textbf{Two-level dynamic load balancing}   &                                                                                                                                   &                                                                                                                          \\
					\begin{tabular}[c]{@{}l@{}}  Thread-level \\ self-scheduling \end{tabular}     & \begin{tabular}[c]{@{}l@{}}STATIC\\ SS, GSS, FSC\tnote{*}, TSS, FAC, RAND\end{tabular}                                                     & \begin{tabular}[c]{@{}l@{}}Static: used as a baseline in this level\\ Dynamic and nonadaptive\end{tabular}                                                     \\ \cline{2-3} 
					\begin{tabular}[c]{@{}l@{}} Process-level\\ self-scheduling \end{tabular}    & \begin{tabular}[c]{@{}l@{}}NODLB\\SS\tnote{*}, FSC\tnote{*}, mFSC, GSS, TSS, FAC, WF\tnote{*}\\ AWF, AWF-B, -C, -D, -E, AF\end{tabular}                    & \begin{tabular}[c]{@{}l@{}}Static: used as a baseline at this level\\ Dynamic and nonadaptive\\ Dynamic and adaptive\end{tabular}                                  \\  \midrule

					\multirow{2}{*}{\textbf{Computing systems}} & \begin{tabular}[c]{@{}l@{}}\alir{miniHPC}\end{tabular}  & \begin{tabular}[c]{@{}l@{}} $20$ Dual socket Intel Broadwell nodes, $10$ cores per socket, $64$ GB RAM per node \\ Two-level nonblocking fat-tree topology with Intel Omni-Path interconnection fabrics\\ Network bandwidth: $100$~Gbit/s, Network latency $100$ nanoseconds \end{tabular}   \\ 	
					
					\cline{2-3}
					& \begin{tabular}[c]{@{}l@{}}\alir{Piz Daint}\end{tabular}  & \begin{tabular}[c]{@{}l@{}} $100$ Cray XC50 single socket Intel Haswell nodes, $12$ cores per socket, $64$ GB RAM per node \\ Dragonfly topology with Cray Aries routing and communications ASIC\\ Network bandwidth: $150$~Gbit/s, Network latency $130$ nanoseconds \end{tabular}   \\

					\bottomrule                                                          
				\end{tabular}
				\begin{tablenotes}
					\item[*] DLS techniques \aliA{implemented in the} \elap{} or \aliA{the} \dlbTool{} libraries but not used in this work \aliA{due to being} unsuitable (SS, WF) or requiring profiling (FSC).
				\end{tablenotes}
			\end{threeparttable}
		\end{adjustbox}
	\end{center}
\end{table}

\subsection{Design of Experiments}
\label{subsec:experiments}
\aliA{Due to the multitude of factors and parameters that need to be considered for the experimental evaluation of the proposed \mbox{two-level} dynamic load balancing approach, we design the experiments as a set of factorial experiments. 
The details used in this design are included in Table~\ref{tbl:design_of_exp}}. 
\subsubsection{Applications}

Mandelbrot is a \mbox{computationally-intensive} application which computes the Mandelbrot set~\cite{mandelbrot1980fractal} and generates its image. 
The application is parallelized such that the calculation of the value at every single pixel of a 2D image is a task, that is performed in parallel. 
To increase the variability between task execution times, the calculation is focused on the center of the image, i.e., the seahorse valley, where the computation is highly \mbox{intensive}.
Mandelbrot is often used to evaluate the performance of dynamic scheduling techniques due to the high variation between its tasks execution times.
Algorithm~\ref{algo:Mandelbrot} shows the calculation of the Mandelbrot set for every pixel to generate a Mandelbrot set image.
The for loop in Line~$2$ is parallelized with DLS techniques and the values of $start$ and $end$ are calculated based on the chunk size obtained by a DLS technique.
The application computes the function $f_c(z) = z^4 + c$ instead of $f_c(z) = z^2 + c$ to increase the number of computations per task (see Lines~10-12). 
Line~9 represents the main source of load imbalance, as the number of repetitions of the calculations between Lines~9 to 14 is irregular.

\begin{algorithm2e}[!h]
	\caption{Mandelbrot set calculation}
	\label{algo:Mandelbrot}
	\SetKwInOut{Input}{Inputs}
	\SetKwInOut{Output}{Output}
	\setcounter{AlgoLine}{0}
	\LinesNumbered
	
	\Input{$W$: \mbox{image width}\\ $K$: \mbox{max iterations}\\ $RM$: real max\\ $Rm$: real min \\ $IM$: image max\\ $Im$:image min \\ $SR$: scale real \\$SI$: scale image\\ $SC$:scale color \\ $data$: pixel information} 
	
	$N = 2$
	
	\tcc{start and end are set by the scheduling techinque according to the chunk size}
	\tcc{calculate pixels in parallel}
	\For{	{\color{blue} $i = start \rightarrow end$ }}
	{ 
		$z.real = z.imag = 0$
		
		$rowID = i/W$
		
		$colID = i mod W$
		
		$c.real = Rm + colID \times SR$
		
		$c.imag = Im + (W - 1 - rowID) \times SI$
		
		$k = 0$
		$lengthsq = 0$
		
		\While{	{\color{blue} $lengthsq < (N \times N)$ }}{
			
			$temp = z.real^4 - 6\times z.imag^2 \times z.real^2  +  z.imag^4 + c.real$
			
			$z.imag = 4 \times z.real^3 \times z.imag -4\times z.real \times z.imag^3 +c.imag$
			
			$z.real = temp$
			
			$lengthsq = z.real^2 + z.imag^2$
			
			$k++$  }
		
		data[i] = $(k-1)\times SC$ }
\end{algorithm2e}

The second application of interest is an application from the computer vision domain, namely the  parallel spin-image algorithm~(PSIA)~\cite{psia}.
PSIA converts a 3D object into a set of 2D descriptors (\mbox{spin-images}).

Algorithm~\ref{algo:PSIA} describes the steps of generating \mbox{spin-images} in PSIA and how it is parallelized.
According to Algorithm~\ref{algo:PSIA}, Lines~9 and 12, the amount of computations to generate \mbox{spin-images} is \mbox{data-dependent} and not identical over all the \mbox{spin-images} generated from the same object. 
This introduces an algorithmic source of load imbalance among the parallel processes generating the \mbox{spin-images}.
The number of \mbox{spin-images} generated by each PE is governed by the \texttt{start} and \texttt{end} variables in Algorithm~\ref{algo:PSIA}, Line~1, which is performed in parallel. 
\begin{algorithm2e}[!h]
	\SetKwInOut{Input}{Inputs}
	\SetKwInOut{Output}{Output}
	\setcounter{AlgoLine}{0}
	\LinesNumbered
	adCalculateSpinImages (W, B, S, OP, M, spinImages, start, end)
	
	\Input{W: \mbox{image width}\\ B: \mbox{bin size}\\ S: \mbox{support angle}\\ \mbox{OP: list of oriented points}\\ \mbox{M: number of oriented points}\\ \mbox{spinImages: list of spin-images to be filled}}
	
	\For{	{\color{blue} imageCounter = start $\rightarrow$ end }}
	{ 
		P = OP[imageCounter]
		
		tempSpinImage[W, W]
		
		init(tempSpinImage)
		
		\For{j = 0 $\rightarrow$ $M$}
		{
			X = OP[j]
			
			$np_i$ = getNormal(P)
			
			$np_j$ = getNormal(X)
			
			\If{	{\color{blue} acos($np_i \cdot np_j$) $\le S$}}
			{
				$k$  =  $\Bigg \lceil$ $\cfrac{W/2 - np_i \cdot (X-P) }{B}$ $\Bigg \rceil$ 
				
				\vspace{0.2cm}
				$l$ =  $\Bigg \lceil$  $\cfrac{ \sqrt{||X-P||^2 - (np_i\cdot(X-P))^2} }{B}$  $\Bigg \rceil$

				\If{ {\color{blue} 0 $\le$ k $\textless$ W and 0 $\le$ l $\textless$ W}}
				{ tempSpinImage[k, l]++
				}
			}
		}
		
		add(spinImages, tempSpinImage)
	}
	\caption{ \mbox{Spin-image} calculation~\cite{Eleliemy:2017b}}
	\label{algo:PSIA}
\end{algorithm2e}

\sphynx{}\footnote{Available at http://astro.physik.unibas.ch/sphynx} is a state-of-the-art production smoothed particle hydrodynamics (SPH) code \cite{cabezon2017sphynx}.
It is also one of the few hydrodynamic codes in the literature that can simulate both Type Ia and \mbox{core-collapse} Supernovas, including nuclear reactions, neutrino transport, and general relativity correction terms.
\sphynx{} is a time-stepping application, parallelized using MPI and OpenMP.
Each time-step consists of several \mbox{computationally-intensive} operations~\cite{cabezon2017sphynx}.
The workflow of \sphynx{} is listed in Algorithm~\ref{algo:sphynx}.

\begin{algorithm2e}[]
	\caption{\sphynx{} Computational Workflow}
	\label{algo:sphynx}
	\setcounter{AlgoLine}{0}
	\LinesNumbered
	
	\For{$l\gets t_{init}$ \KwTo $t_{final}$}
	{
		1. Build tree
		
		2. Find neighbors
		
		\hspace{0.75em} 2.1 Collective communication (number of neighbors) 
		
		3. Density \& grad-h calculations
		
		\hspace{0.75em} 3.1 Collective communication (density \& grad-h)
		
		4. IAD calculations
		
		\hspace{0.75em} 4.1 Collective communication (IAD terms)
		
		5. EOS \& $\nabla\textbf{v}$ calculations
		
		\hspace{0.75em} 5.1 Collective communication ($\nabla\cdot\textbf{v}$ \& $\nabla\times\textbf{v}$)
		
		6. Momentum \& energy calculations
		
		\hspace{0.75em} 6.1 Collective communication ($\nabla P$ \& $du/dt$)
		
		7. Gravity calculations
		
		\hspace{0.75em} 7.1 Collective communication (gravitational force and potential)
		
		8. Update velocities, position, and energy
		
		9. Time-step evaluation
		
		10. Verification via conservation laws
		
	}
\end{algorithm2e}

The performance of \sphynx{} is studied {for} two simulation test-cases. 
{The \textit{stellar collision} test simulates the \mbox{head-on} impact of two \mbox{Sun-like} stars. 
This simulation has two independent gravitating bodies and, therefore, the particle distribution is highly asymmetric. 
Second, the \textit{Evrard collapse} is a common {test} used to {examine} the coupling between hydrodynamics and \mbox{self-gravity} in astrophysical codes. 
It simulates the collapse of an unstable cloud of gas and the formation of the subsequent shock-wave.}
This test is also studied on a large domain size (10 million particles) to explore its performance at large scale.
These {two} \mbox{test-cases} offer a wide range of problem sizes, defined in the number of particles in the system and different particle distributions, that represent different load balancing challenges.

{\figurename{~\ref{fig:load_imbalance}} and \figurename{~\ref{fig:load_imbalance2}} show the load imbalance in the $3$ applications of interest with no load balancing at the \tl{} and \pl{} executing on miniHPC\footnote{
miniHPC is a fully controlled {research and teaching} HPC cluster at the Department of Mathematics and Computer Science at the University of Basel, Switzerland, https://hpc.dmi.unibas.ch/HPC/miniHPC.html.}
This load imbalance manifests as {overhead (i.e., waiting time)} as depicted in \figurename{~\ref{fig:load_imbalance}} at the \mbox{thread-level}~(yellow regions) and at the \mbox{process-level}~(red regions). 
{As calculating gravity is the most \mbox{time-consuming} computational step and also the most load imbalanced, this work focuses on improving the gravity calculation step of \sphynx{}.}
{To obtain representative performance measurements, the \mbox{two-level} load balancing is tested in the middle of the simulation time~(\mbox{time-step} $6900$) for the \textit{stellar collision} \mbox{test-case}.
For the Evrard collapse \mbox{test-case}, all DLS combinations at the two levels are tested at multiple snapshots of the simulation as listed in Table~\ref{tbl:design_of_exp}.
After testing all DLS combinations at different simulation stages, the identified best \mbox{two-level} DLS combination is used to execute the full simulation to measure the achieved overall performance improvement for a full SPH simulation.}
It {is} worth {noting} that the observed load imbalance in~\figurename{~\ref{fig:load_imbalance}} for a single \mbox{time-step} of \sphynx{} can {repeatedly} be observed through the execution of the full SPH simulation, which typically {requires} $10^5$ to $10^6$ \mbox{time-steps}. 

\begin{figure}[p!]
		\centering
		\subfloat[Mandelbrot, $40$ processes, $10$ threads (miniHPC)]{%
			\includegraphics[clip, trim=0cm 0cm 0cm 0cm,scale=0.168]{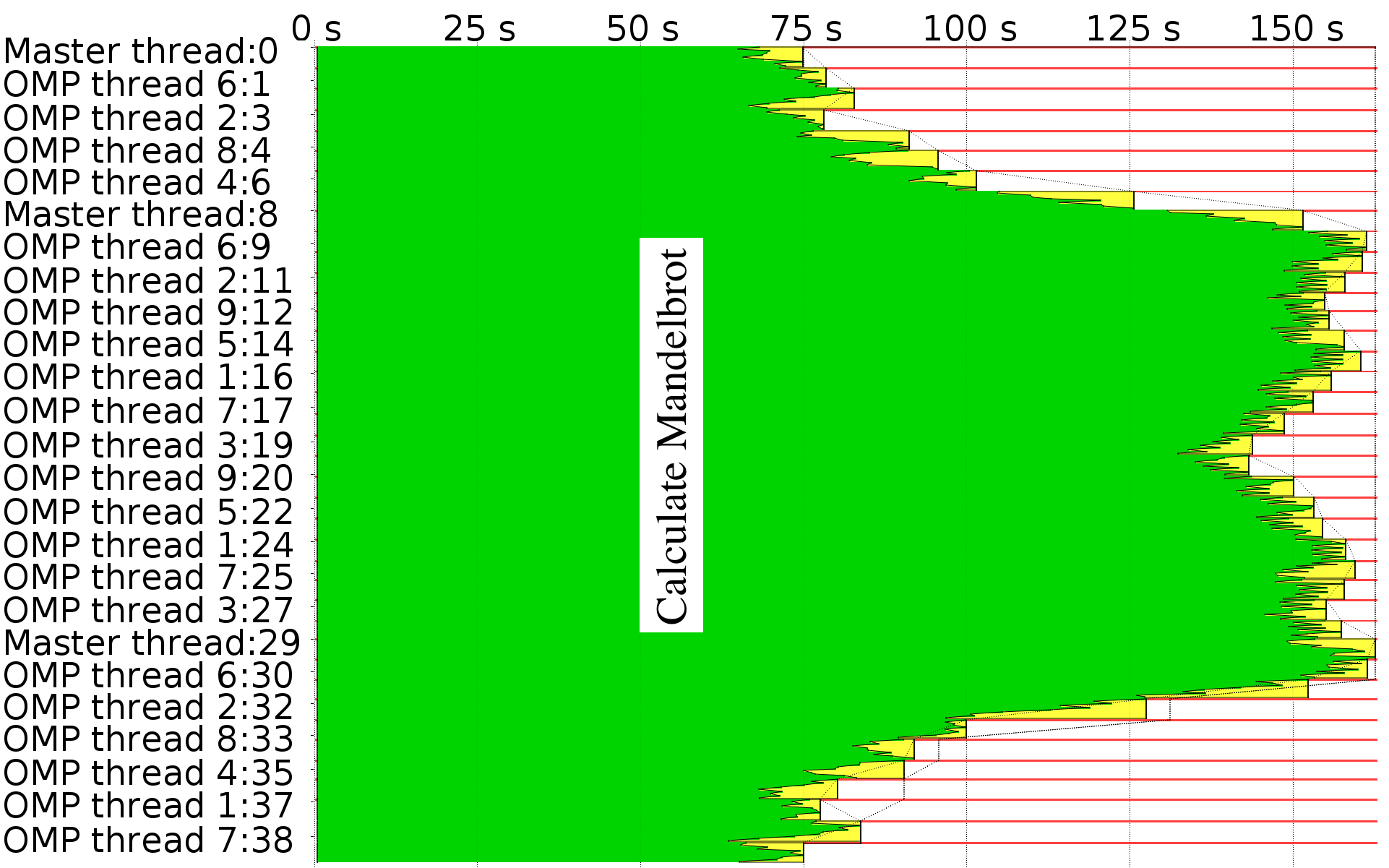}
			\label{subfig:Mandelbrot_imbalance}%
		} \hspace{0.0cm} 
	\subfloat[PSIA, $40$ processes, $10$ threads (miniHPC)]{%
		\includegraphics[clip, trim=0cm 0cm 0cm 0cm, scale=0.168]{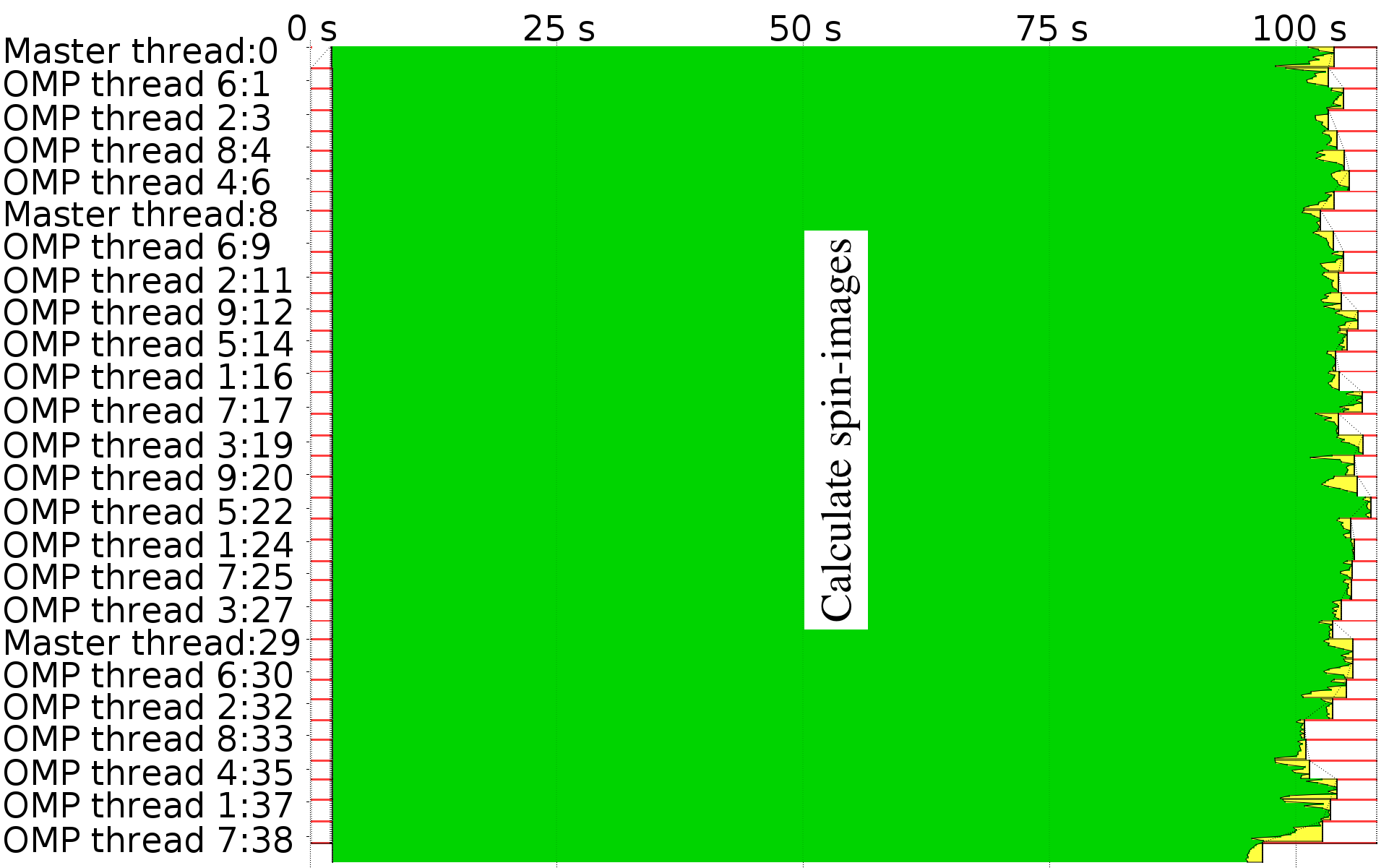}%
		\label{subfig:PSIA_imbalance}%
	}  \\
  \subfloat[Stellar collision time-step $6916$, \sphynx{}, $40$ processes, $10$ threads (miniHPC)]{%
	\includegraphics[clip, trim=0cm 0cm 0cm 0cm, scale=0.168]{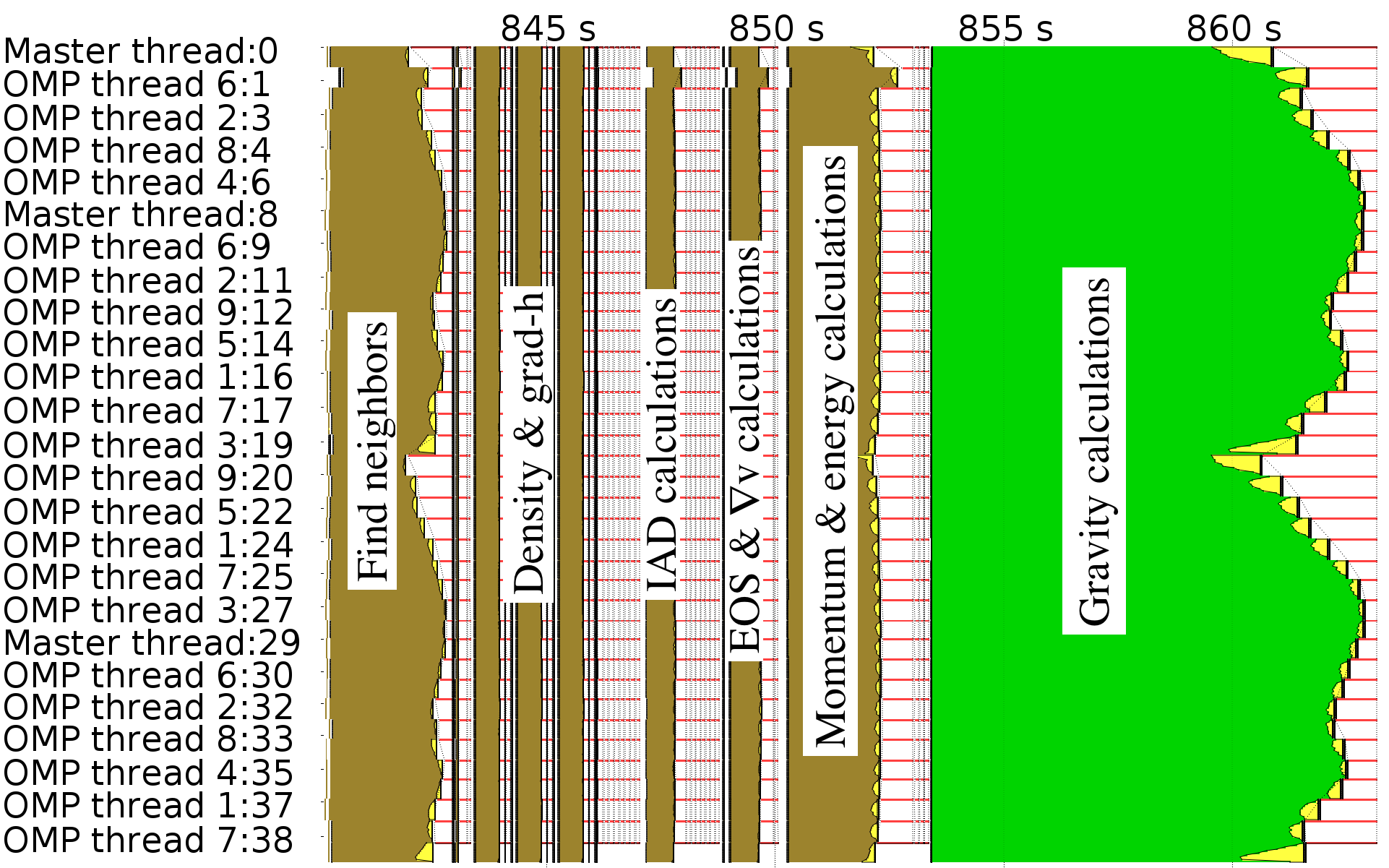}
	\label{subfig:collision_imbalance}%
} \hspace{0.0cm} 
	\subfloat[Evrard, $1M$ particles, time-step $116$, \sphynx{}, $12$ processes, $10$ threads (miniHPC)]{%
	\includegraphics[clip, trim=0cm 0cm 0cm 0cm, scale=0.168]{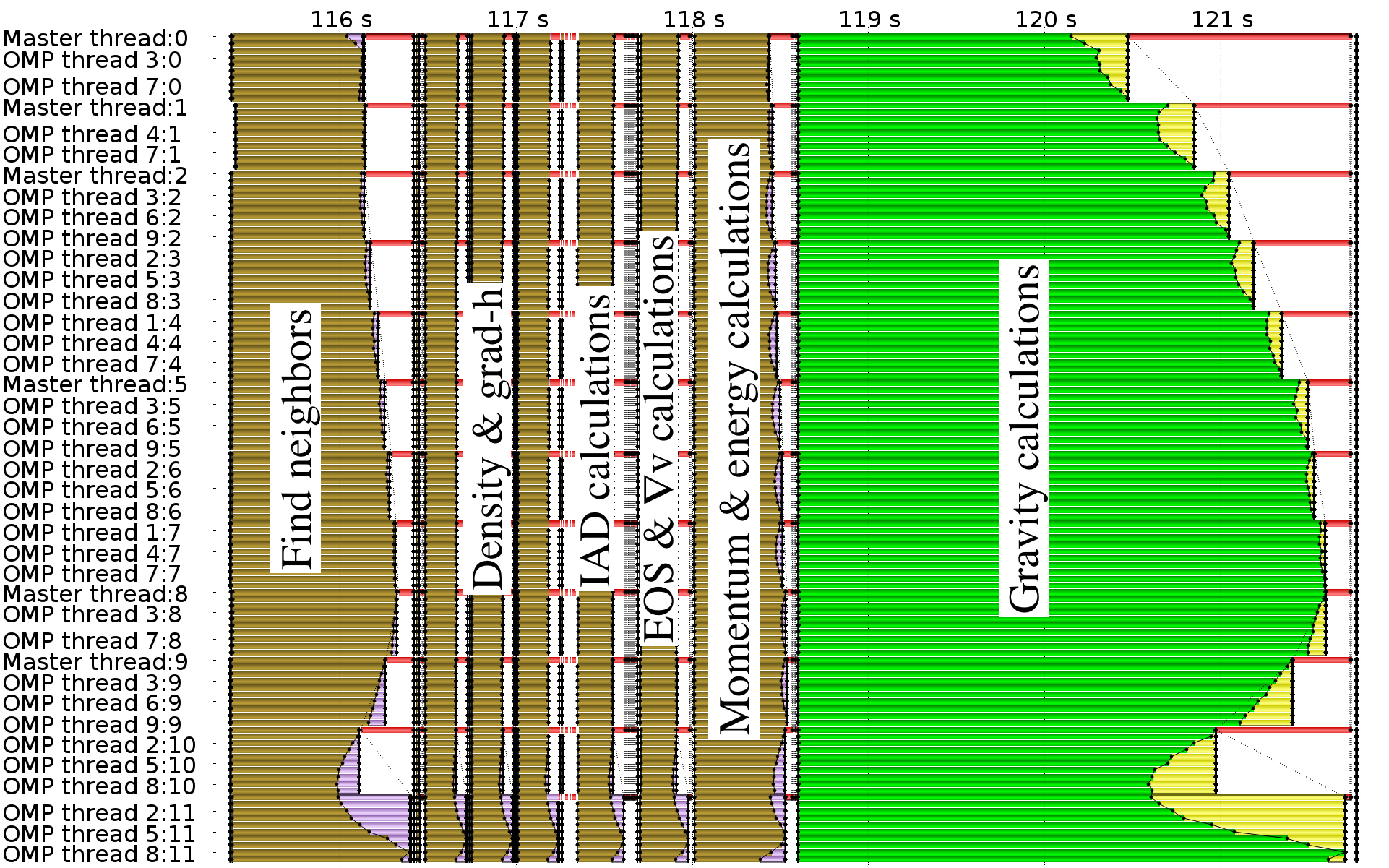}
	\label{subfig:evrard_100_imbalance}%
}\\
	\subfloat[Evrard, $1M$ particles, time-step $516$, \sphynx{}, $12$ processes, $10$ threads (miniHPC)]{%
	\includegraphics[clip, trim=0cm 0cm 0cm 0cm, scale=0.168]{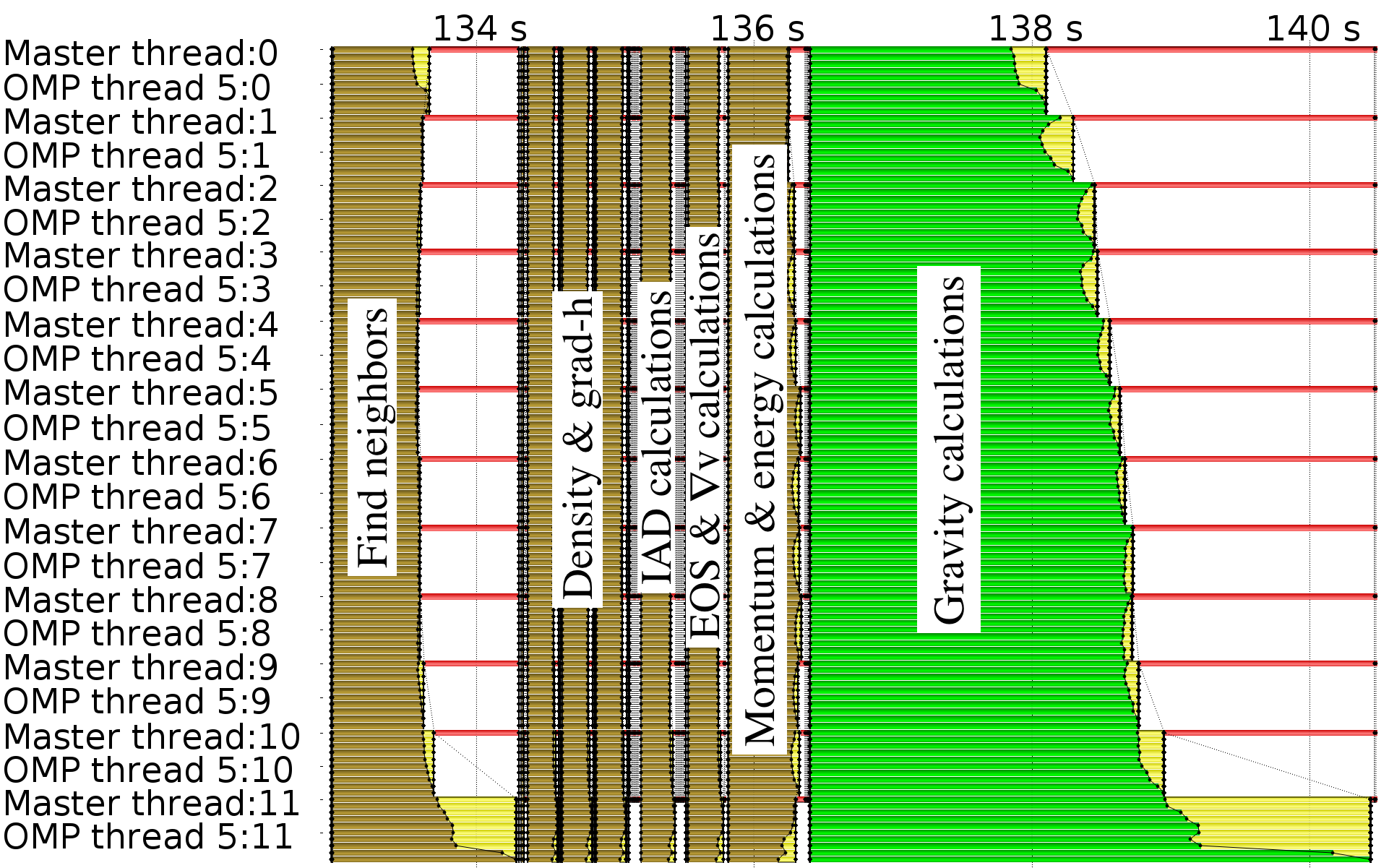}
	\label{subfig:evrard_500_imbalance}%
}\hspace{0.0cm} 
	\subfloat[Evrard, $1M$ particles, time-step $1016$, \sphynx{}, $12$ processes, $10$ threads (miniHPC)]{%
	\includegraphics[clip, trim=0cm 0cm 0cm 0cm, scale=0.168]{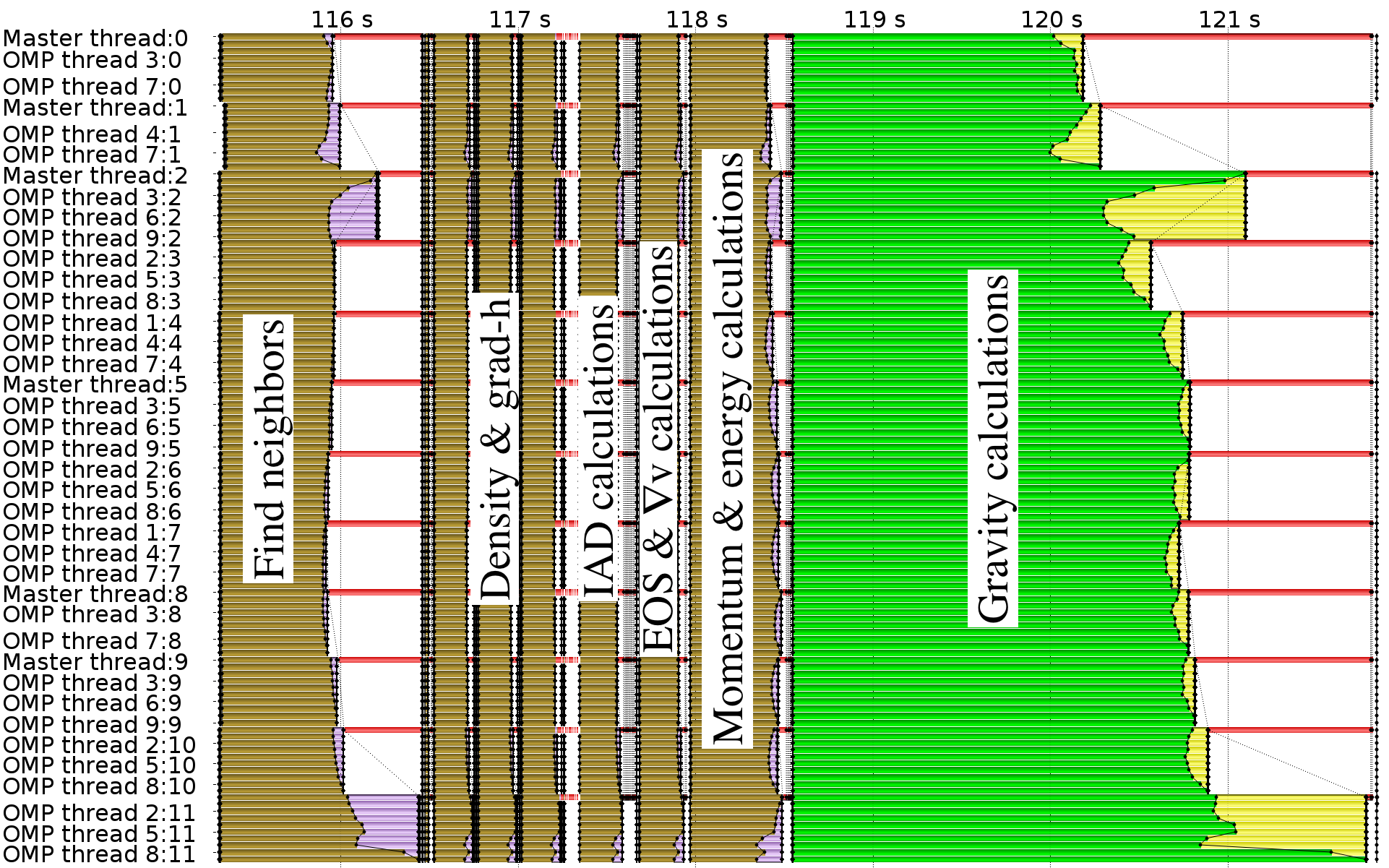}
	\label{subfig:evrard_1000_imbalance}%
}\\
	\caption{Impact of two-level load imbalance at \tl{} and \pl{} in the three scientific applications. Idle time due to load imbalance is shown in yellow at the \tl{} and in red at the \pl{} level.
	We apply DLS for each application to the load imbalance in their respective green regions.}
	\label{fig:load_imbalance}
\end{figure}
\clearpage
\begin{figure}[p!]
		\centering
		\subfloat[Evrard, $1M$ particles, time-step $1716$, \sphynx{}, $12$ processes, $10$ threads (miniHPC)]{%
			\includegraphics[clip, trim=0cm 0cm 0cm 0cm, scale=0.168]{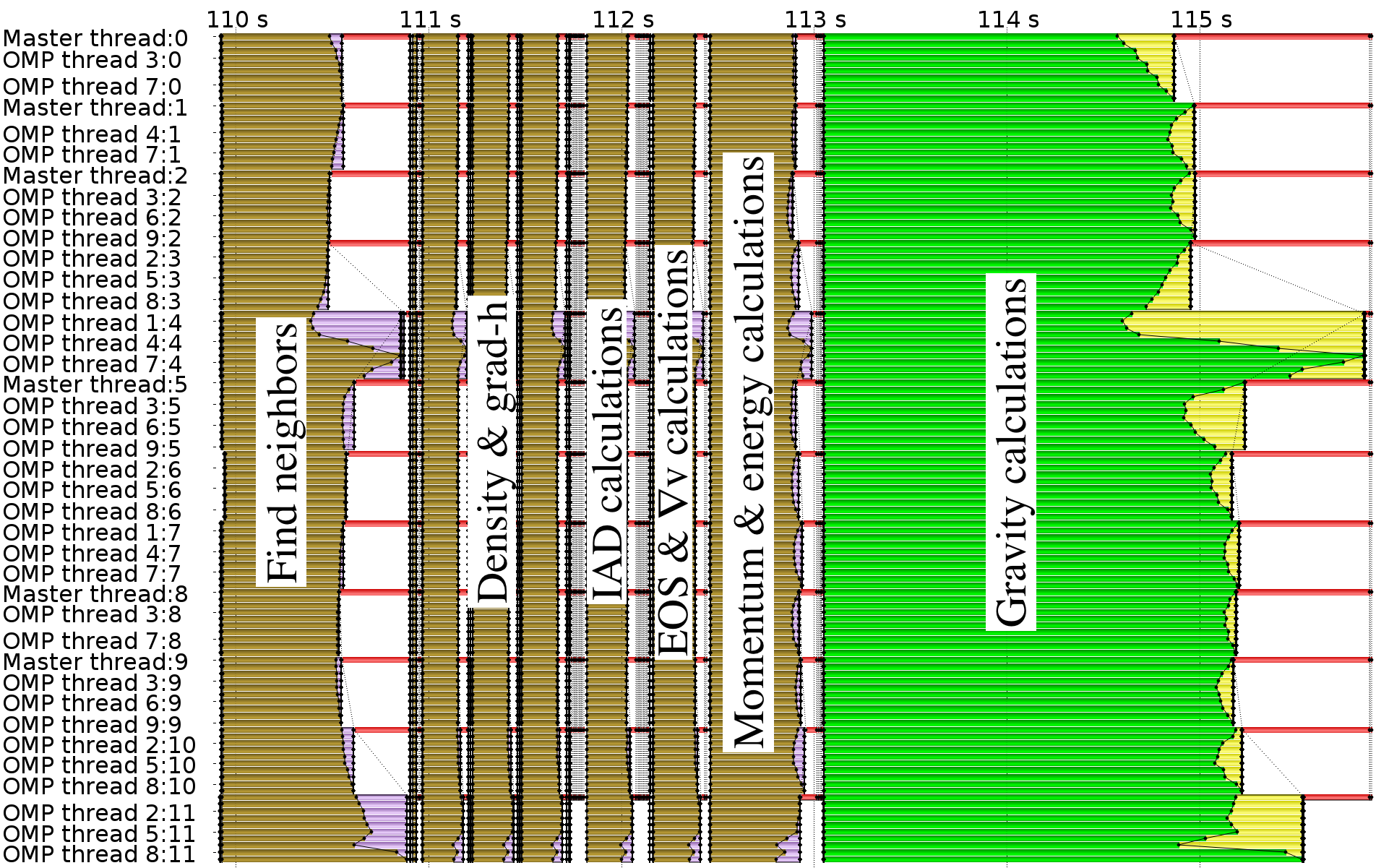}
			\label{subfig:evrard_1700_imbalance}%
		}\hspace{0.0cm} 
		\subfloat[Evrard, $1M$ particles, time-step $2016$, \sphynx{}, $12$ processes, $10$ threads (miniHPC)]{%
			\includegraphics[clip, trim=0cm 0cm 0cm 0cm, scale=0.168]{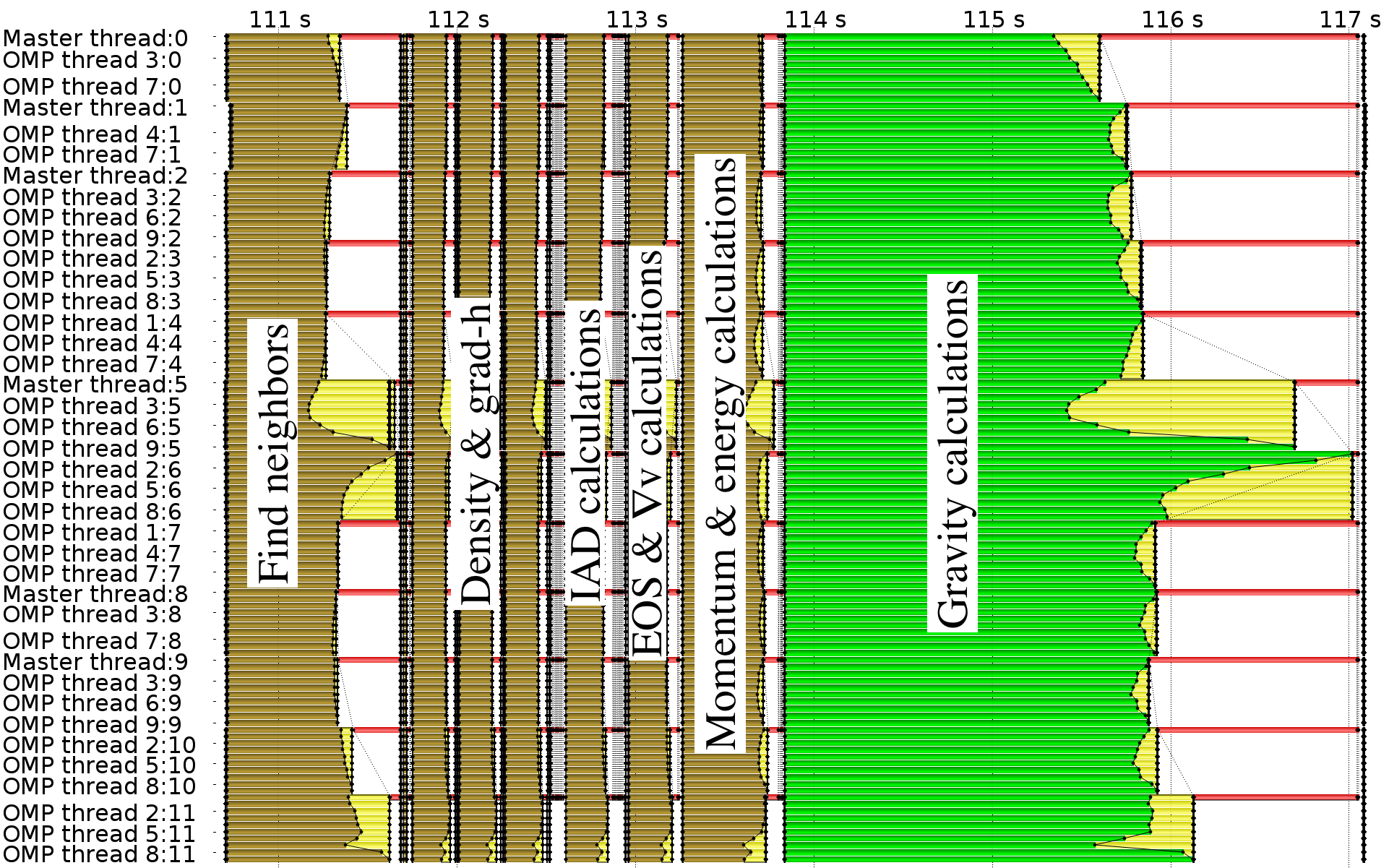}
			\label{subfig:evrard_2000_imbalance}%
		}\\
		\subfloat[Evrard, $1M$ particles, time-step $2316$, \sphynx{}, $12$ processes, $10$ threads (miniHPC)]{%
		\includegraphics[clip, trim=0cm 0cm 0cm 0cm, scale=0.168]{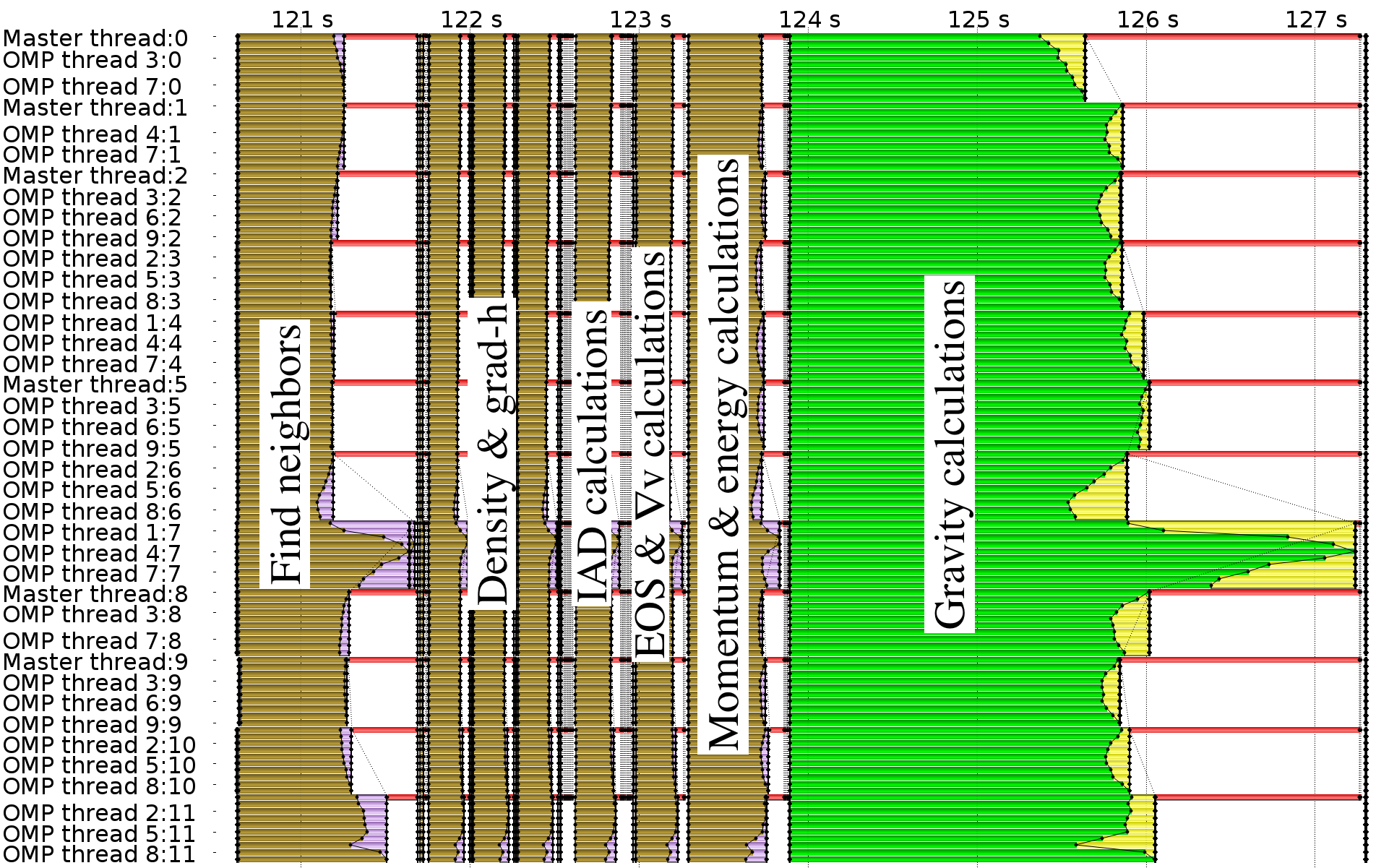}
		\label{subfig:evrard_2300_imbalance}%
	}\hspace{0.0cm} 
	\subfloat[Evrard, $1M$ particles, time-step $2516$, \sphynx{}, $12$ processes, $10$ threads (miniHPC)]{%
		\includegraphics[clip, trim=0cm 0cm 0cm 0cm, scale=0.168]{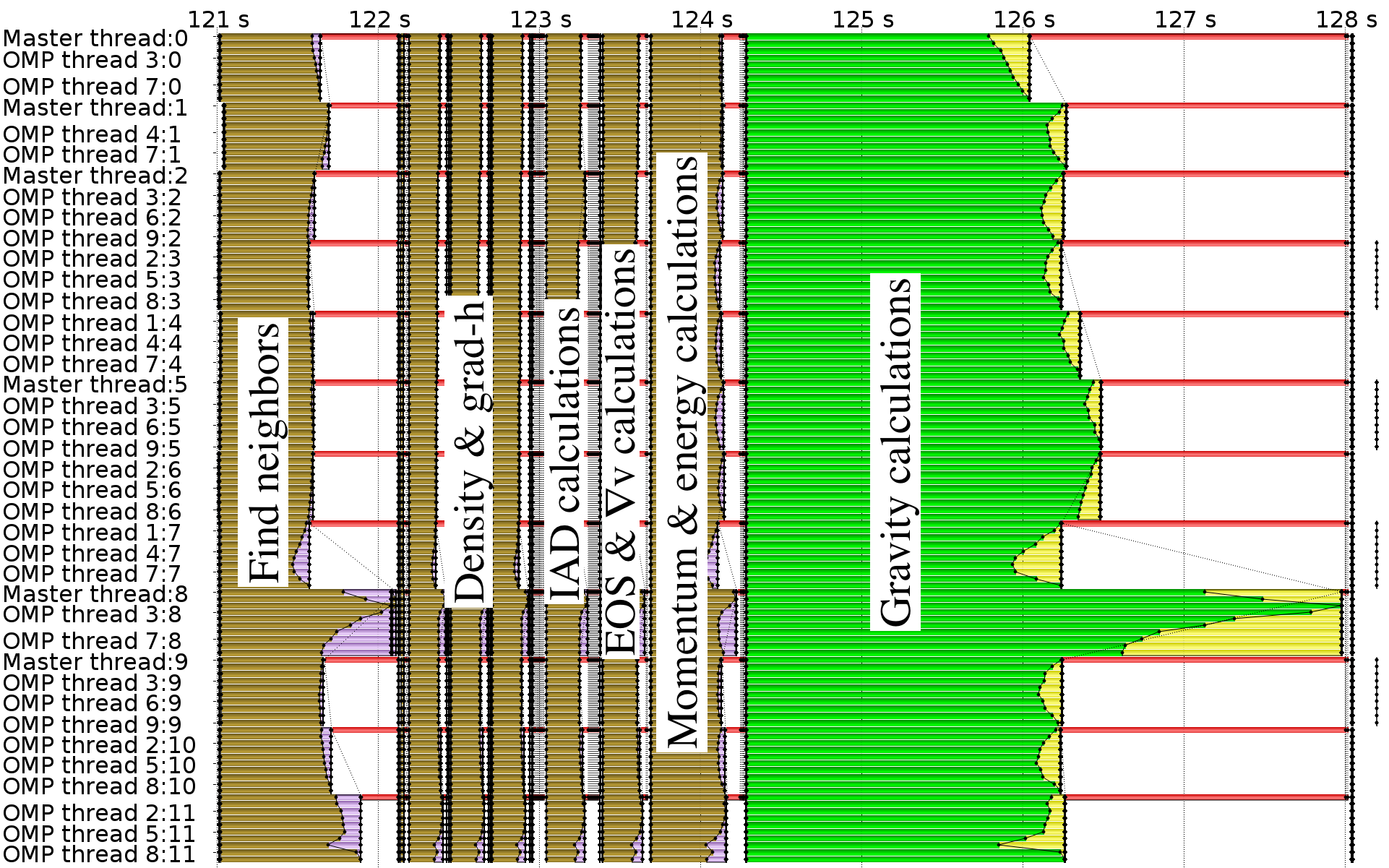}
		\label{subfig:evrard_2500_imbalance}%
	}\\
	\subfloat[Evrard, $1M$ particles, time-step $2816$, \sphynx{}, $12$ processes, $10$ threads (miniHPC)]{%
	\includegraphics[clip, trim=0cm 0cm 0cm 0cm, scale=0.168]{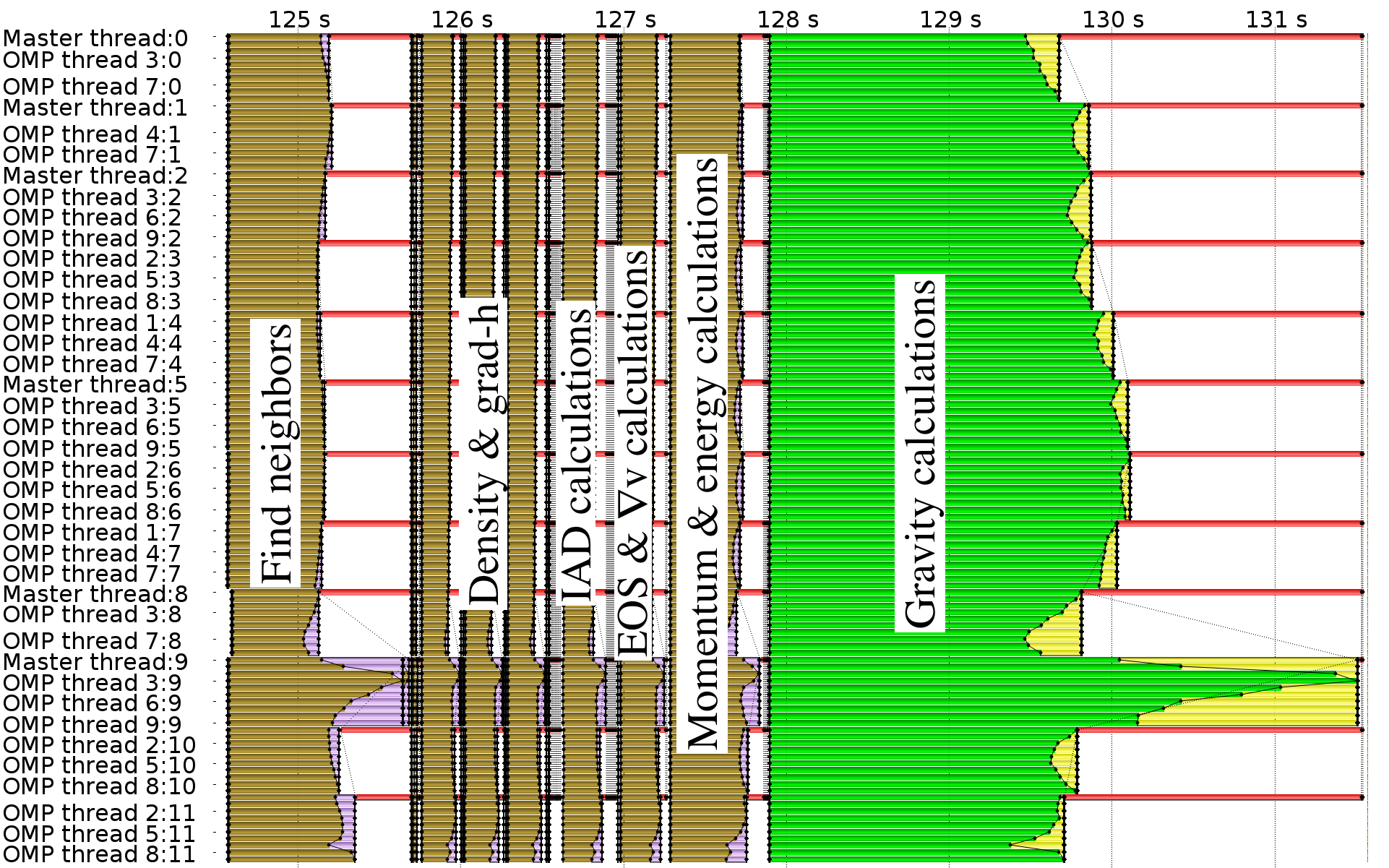}
	\label{subfig:evrard_2800_imbalance}%
}\\
		\caption{Impact of two-level load imbalance at \tl{} and \pl{} in the Evrard collapse test-case, $1M$ particles, with \sphynx{}. Idle time due to load imbalance is shown in yellow at the \tl{} and in red at the \pl{} level. We apply DLS to the load imbalance in their green region corresponding to gravity calculations.}
		\label{fig:load_imbalance2}
\end{figure}
\clearpage

\subsubsection{Two-level dynamic load balancing}
Six loop scheduling techniques \aliA{are considered} at the \tl{} via the \elap{} OpenMP library and eleven~loop scheduling techniques at the \pl{} via the \dlbTool{}, \ali{yielding} a combination of $6 \times 11 = 66$~experiments per application or \mbox{test-case}.
\aliA{FSC technique is not considered in this work neither at the \tl{} nor the \pl{} as it requires the profiling of the application to estimate the standard deviation of task execution times $\sigma$ and the scheduling overhead $h$. 
Application performance with FSC is significantly influenced by the provided $\sigma$ and $h$ values.

The SS technique is not considered at the \pl{}, as it assigns a single task to a requesting process, which waste the \tl{} parallelism as only one thread is used.
WF technique is not used in this work as well, as it is designed for heterogeneous computing systems, which is not the case for miniHPC.
AWF technique is only used with \sphynx{} as it learns PE relative performance weights from previous \mbox{time-steps}.}
\texttt{NODLB} and STATIC \ali{denote the scenario where} \aliA{application tasks are statically and equally divided among the processes or threads, respectively.} 

\aliA{A minimum chunk size is specified at the \pl{} to avoid processes being assigned a very small chunk of tasks towards the end of the execution, which could not contain enough work to distribute to threads within a process and increase the scheduling rounds and consequently the overall scheduling overhead.
The minimum chunk size at the \pl{} is set to half the chunk size of mFSC technique, that is $532$, $700$, $ 7278$, $2549$, for Mandelbrot, PSIA, \sphynx{} with stellar collision test, and \sphynx{} with Evrard collapse test, respectively.
At the \tl{} the minimum chunk size is set to $1$ (OpenMP default) as the scheduling overhead is small and to achieve the best possible load balance.}

\subsubsection{Computing systems}
The \aliA{proposed} is tested \ali{on} miniHPC and Piz~Daint\footnote{Piz Daint is a Cray XC50 supercomputer system, the Swiss National Supercomputing Center~(CSCS), https://www.cscs.ch/computers/dismissed/piz-daint-piz-dora/.}.
\aliA{Each MPI rank is pinned to a CPU socket of miniHPC, to improve the data locality among threads within an MPI rank.
The number of threads per MPI rank is set to be equal to the number of cores per socket. 
Therefore, each compute node of miniHPC executes two MPI ranks, one per socket, with $10$ OpenMP threads within each MPI rank.}
In addition, we use Piz Daint to run large scale experiments with $100$ nodes and $12$ threads per node for the execution of Evrard collapse \mbox{test-case} with $10$ million particles.
In all experiments, a rank is pinned to a processor socket within a compute node, i.e., two ranks per node on miniHPC and one rank per node on Piz~Daint. 

\subsection{Experimental Results}
The performance results of the three scientific  applications of interest are depicted in \figurename{~\ref{fig:heat_map}}, \figurename{~\ref{fig:heat_map2}} and \figurename{~\ref{fig:per_level_performance}}.
\ali{The figures show the parallel execution time of Mandelbrot and PSIA and the computing time of the \mbox{gravity} per \mbox{time-step} for \sphynx{}.}
\figurename{~\ref{fig:heat_map}} shows the \ali{performance} percent improvement with DLS techniques normalized to not using any dynamic load balancing mechanism \ali{in any of the \tl{} or the \pl{}}~(NODLB\_STATIC).

Using \ali{two-level} load balancing improved the performance of Mandelbrot up to $21\%$ 
as shown in \figurename{~\ref{subfig:Mandelbrot_heatmap}} with \ali{TSS} at the \pl{} and \ali{SS} at the \tl{}.
The performance improvement is much lower in PSIA than in Mandelbrot as PSIA is mildly imbalanced as shown in~\figurename{~\ref{subfig:PSIA_imbalance}}. 
Two-level load balancing improved the performance of gravity calculations in \sphynx{} also by $11\%$ for stellar collision \mbox{test-case} in~\figurename{~\ref{subfig:collision_heatmap}} with \mbox{AWF-C} at the \pl{} and FAC at the \tl{} and by $43\%$ for Evrard collapse \mbox{test-case}, $1M$ particles, in~\figurename{~\ref{subfig:evrard_500_heatmap}} with GSS at the \pl{} and FAC at the \tl{}.

The best two-level DLS combination for the gravity part of Evrard collapse, namely GSS+FAC, also holds when we increase problem size to $10M$ particles.
GSS+FAC is the most efficient combination at the two levels on $100$ nodes and $12$ threads per node on Piz Daint and improves \sphynx{} performance (gravity part) by $11\%$~(\figurename{~\ref{subfig:evrard_1_heat_map}}).
The relatively low performance improvement for the large-scale experiments is attributed to the fact that they are performed from the beginning of the astrophysical simulation, i.e., \mbox{time-step}~1, where computation and load imbalance are less intensive.
The RAND technique is excluded from these large scale experiments due to its poor performance on the Evrard collapse test with $1M$ particles.
We plan to perform additional experiments (snapshots into the middle of the simulation and full simulation) for the Evrard collapse (gravity part and entire application) with $10M$ particles on Piz Daint similar to the Evrard collapse with $1M$ particles on miniHPC. 

In general, FAC result in the best performance at the \tl{} while GSS and AWF variants result in the best performance at the \pl{}.
SS results in poor performance for \sphynx{} at the \tl{} due to the fine granularity of its tasks ($240$~us on average).
At the \pl{}, the AF technique performs poorly for the experiments conducted in this work.
This can be attributed \ali{in part} to its large overhead and the lack of high variability (in application and computing system) to hide this overhead and benefit from AF.
Specifically, \ali{the} AF technique is designed for highly irregular workloads that execute in stochastic environments. 
The experiments conducted in this work have insufficient variability to justify the adjustments of AF.

\figurename{~\ref{fig:per_level_performance}} shows the best average parallel execution time over $20$ repetitions (Mandelbrot and PSIA) or \mbox{time-step} (\sphynx{}) when: 
(1)~Using no load balancing \ali{at any of the} \mbox{two levels};
(2)~Using the best \ali{DLS technique} \ali{only} at the \tl{};
(3)~Using the best \ali{DLS technique} \ali{only} at the \pl{};
(4)~Using the best available combination of \ali{DLS} techniques at the \tl{} and the \pl{}.
The results in \figurename{~\ref{fig:per_level_performance}} show the benefits of two-level load balancing versus \ali{only} one-level as conceptually illustrated in \figurename{~\ref{fig:two-level}}.
The results show that certain performance gains are achieved by \mbox{single-level} load balancing (either \tl{} or \pl{} middle boxes) as predicted by~\figurename{~\ref{fig:two-level}}, however, the best performance is always achieved by \mbox{two-level} dynamic load balancing.

GSS and FAC was identified as the best combination of DLS techniques at the \pl{} and the \tl{}, respectively, by testing all $66$ DLS combinations at the two levels for Evrard collapse test-case, $1M$ particles, at different stages of the simulation at \mbox{time-steps} $100$, $500$, $1000$, $1700$, $2000$, $2300$, $2500$, and $2800$.

\figurename{~\ref{fig:full_simulation}} shows the parallel execution time per \mbox{time-step} for the full simulation of Evrard collapse with $1M$ particles on miniHPC.
The results show that \mbox{time-step} execution time with \mbox{two-level} dynamic load balancing is always better than the baseline with no load balancing at the two levels and lead to an overall application performance improvement of $15\%$.

\begin{figure}[p!]
	\centering
	\subfloat[Mandelbrot, $40$ processes, $10$ threads (miniHPC)]{%
		\includegraphics[clip, trim=0cm 0cm 0cm 0cm,scale=0.43 ]{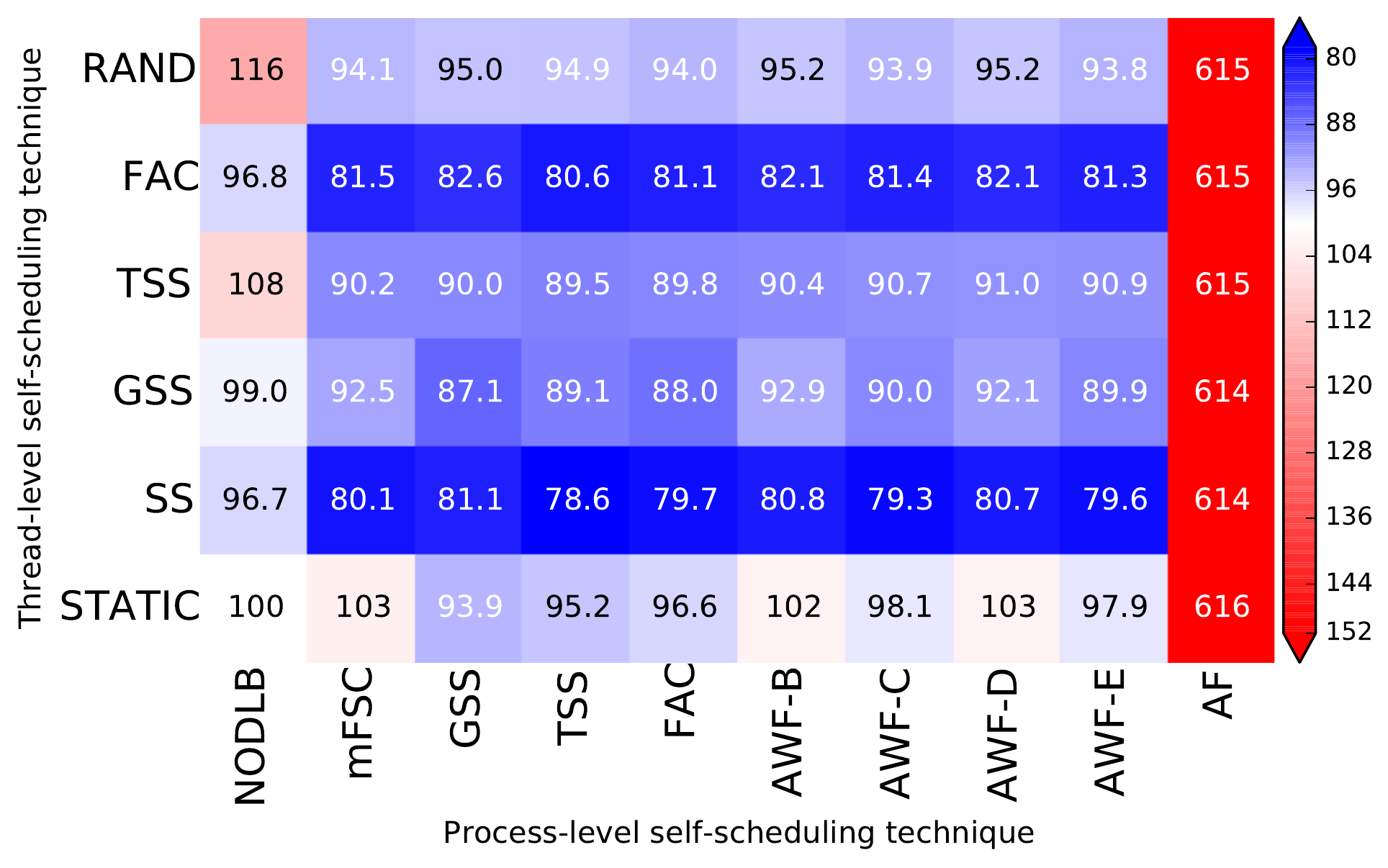}
		\label{subfig:Mandelbrot_heatmap}%
	} \hspace{0cm} 
	\subfloat[PSIA, $40$ processes, $10$ threads (miniHPC)]{%
		\includegraphics[clip, trim=0cm 0cm 0cm 0cm, scale=0.43]{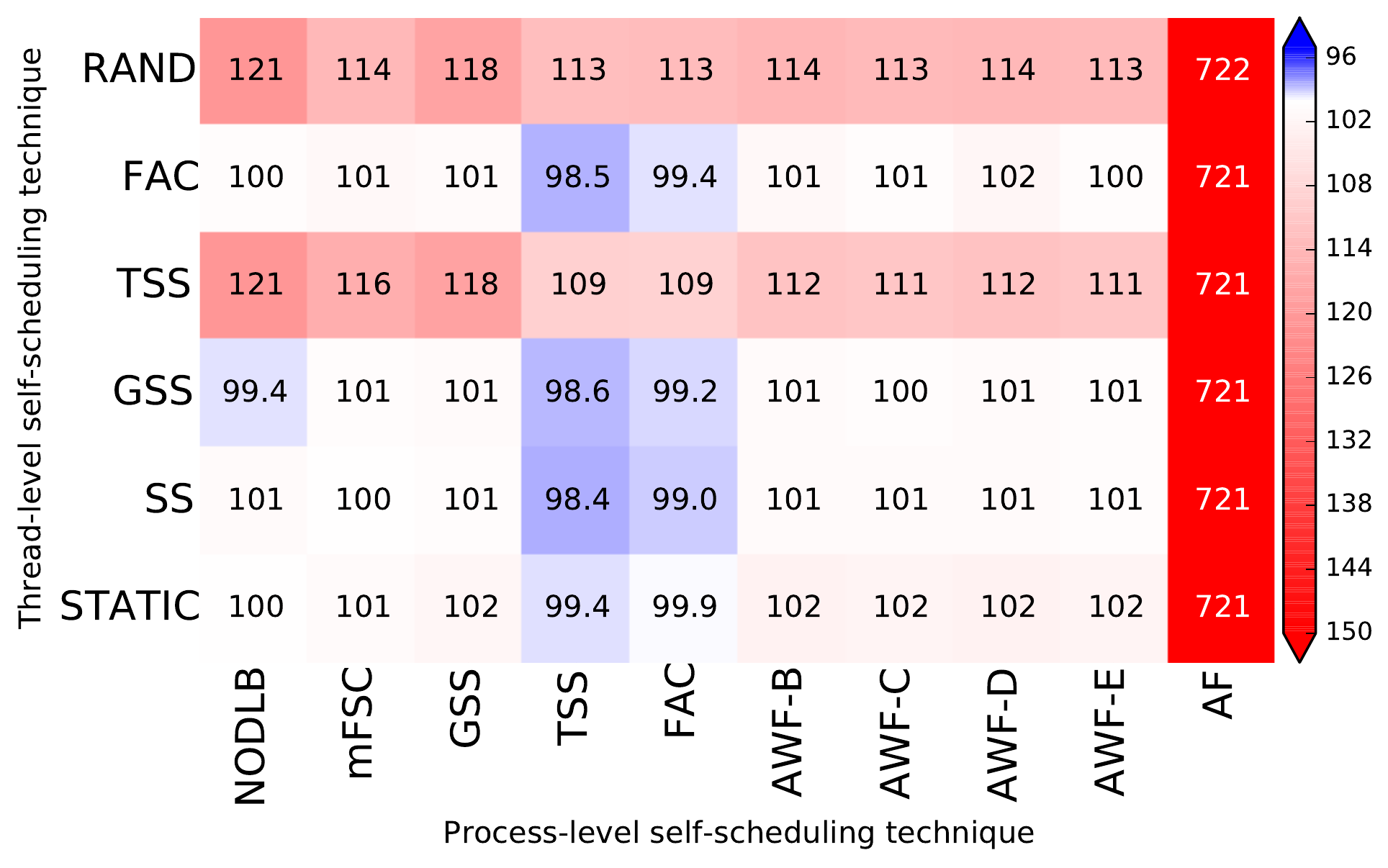}%
		\label{subfig:PSIA_heatmap}%
	} \\
	\subfloat[Stellar collision time-step 6900-6920, $40$ processes, $10$ threads (miniHPC)]{%
		\includegraphics[clip, trim=0cm 0cm 0cm 0cm,scale=0.43]{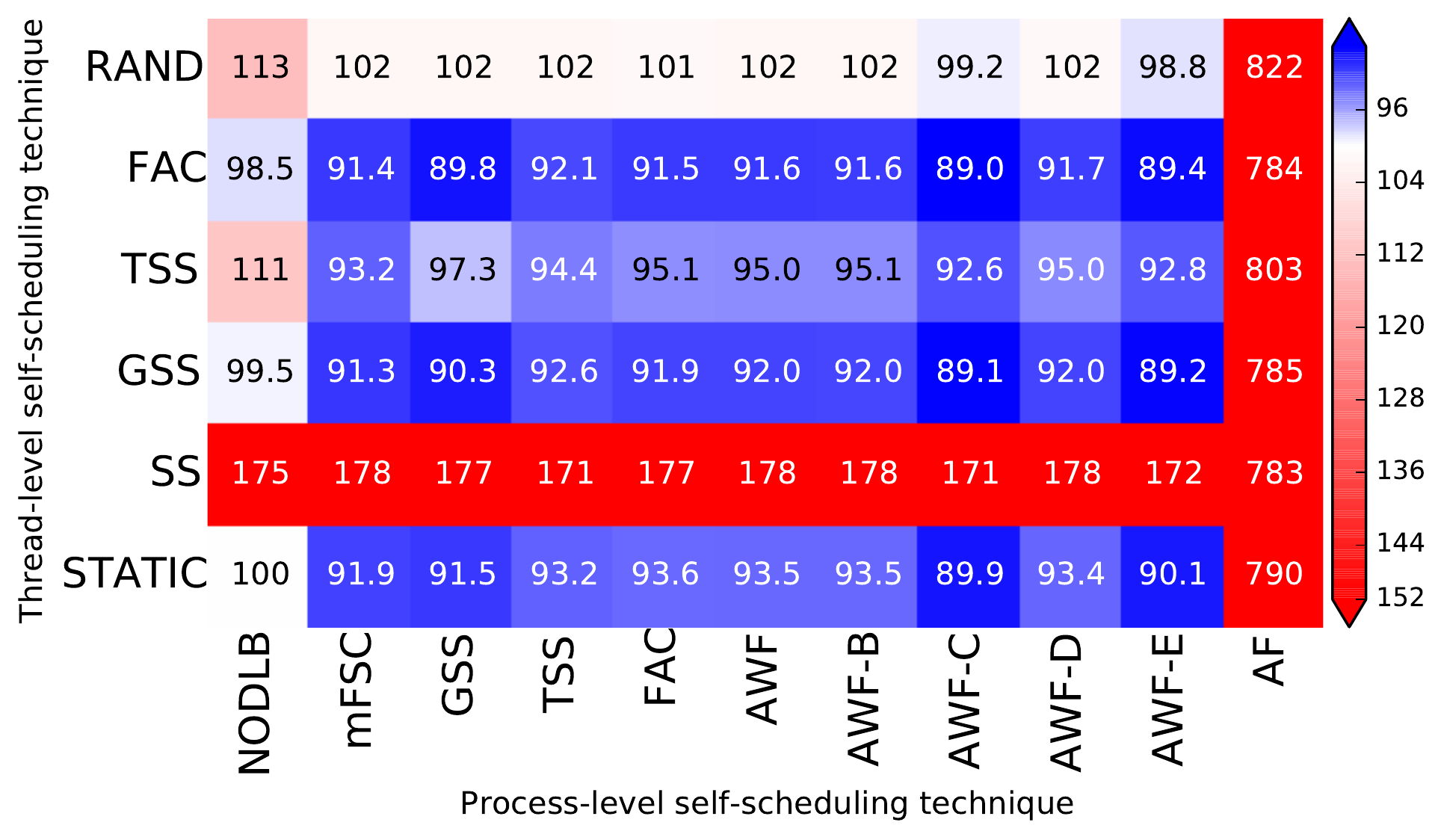}
		\label{subfig:collision_heatmap}%
	} \hspace{0cm}
	\subfloat[Evrard collapse, $1M$ particles, time-step 100-120, $12$ processes, $10$ threads (miniHPC)]{%
		\includegraphics[clip, trim=0cm 0cm 0cm 0cm, scale=0.43]{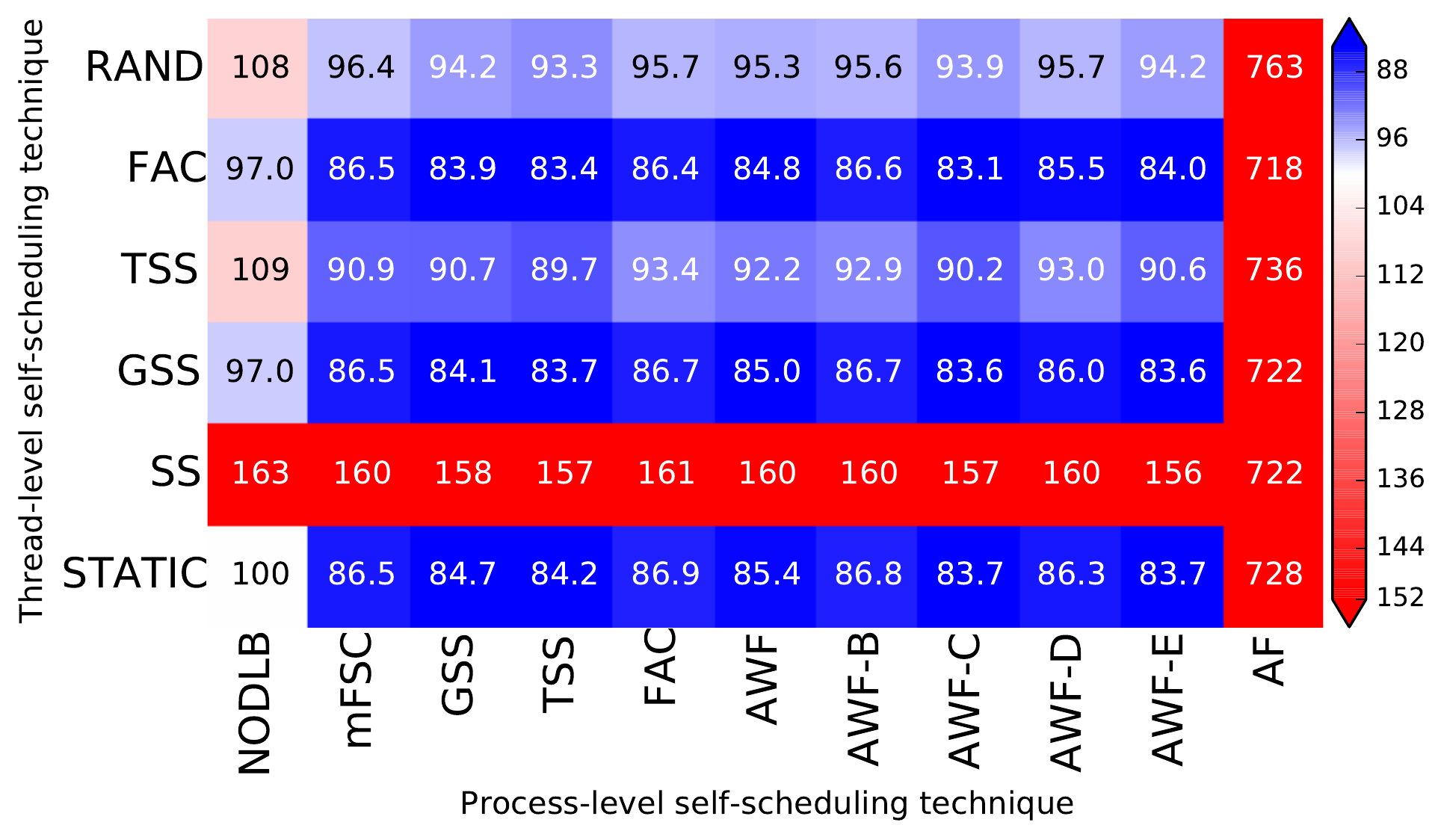}%
		\label{subfig:evrard_100_heatmap}%
	} \\
	\subfloat[Evrard collapse, $1M$ particles, time-step 500-520, $12$ processes, $10$ threads (miniHPC)]{%
		\includegraphics[clip, trim=0cm 0cm 0cm 0cm,scale=0.43]{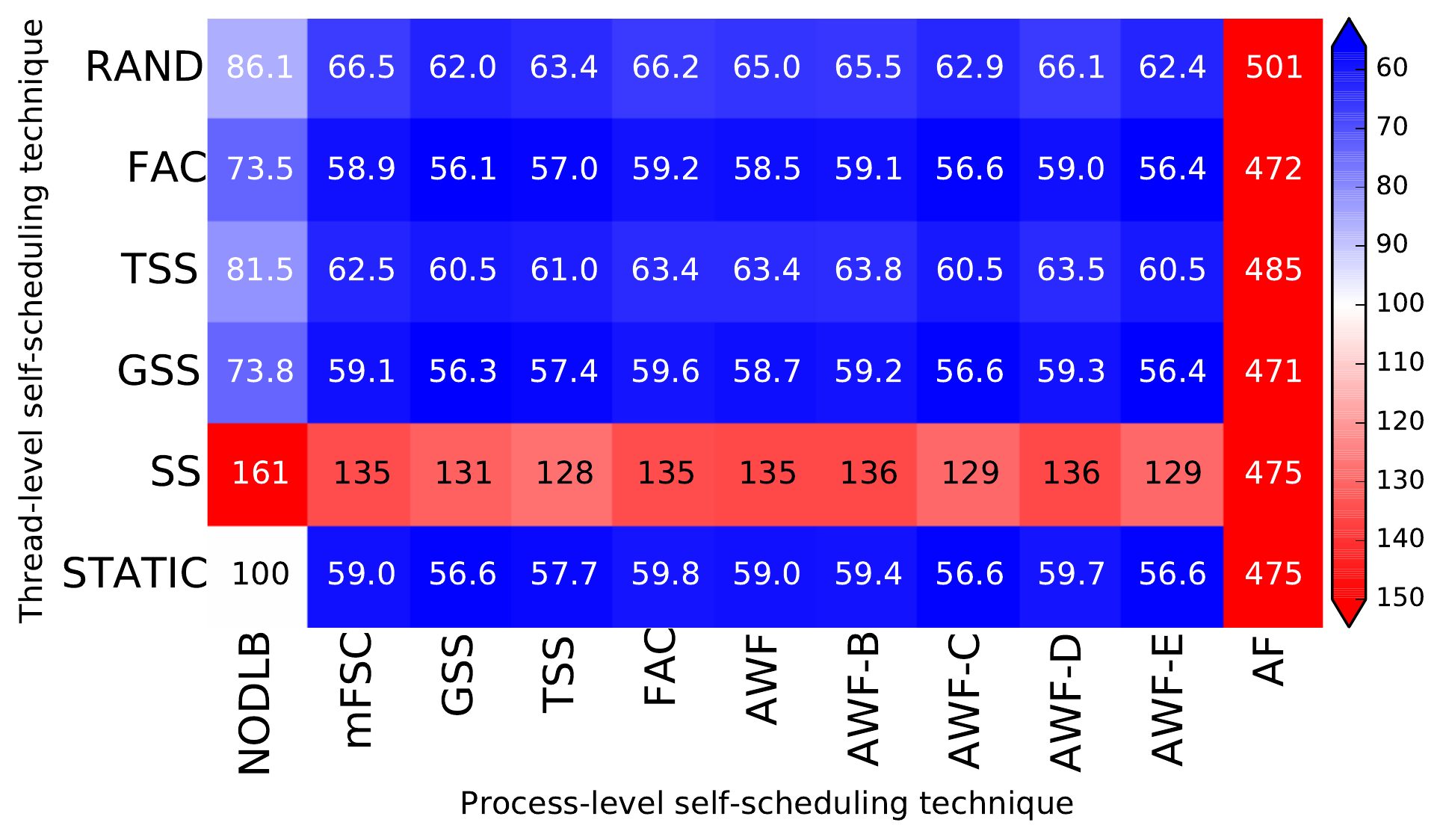}
		\label{subfig:evrard_500_heatmap}%
	} \hspace{0cm}
	\subfloat[Evrard collapse, $1M$ particles, time-step 1000-1020, $12$ processes, $10$ threads (miniHPC)]{%
		\includegraphics[clip, trim=0cm 0cm 0cm 0cm, scale=0.43]{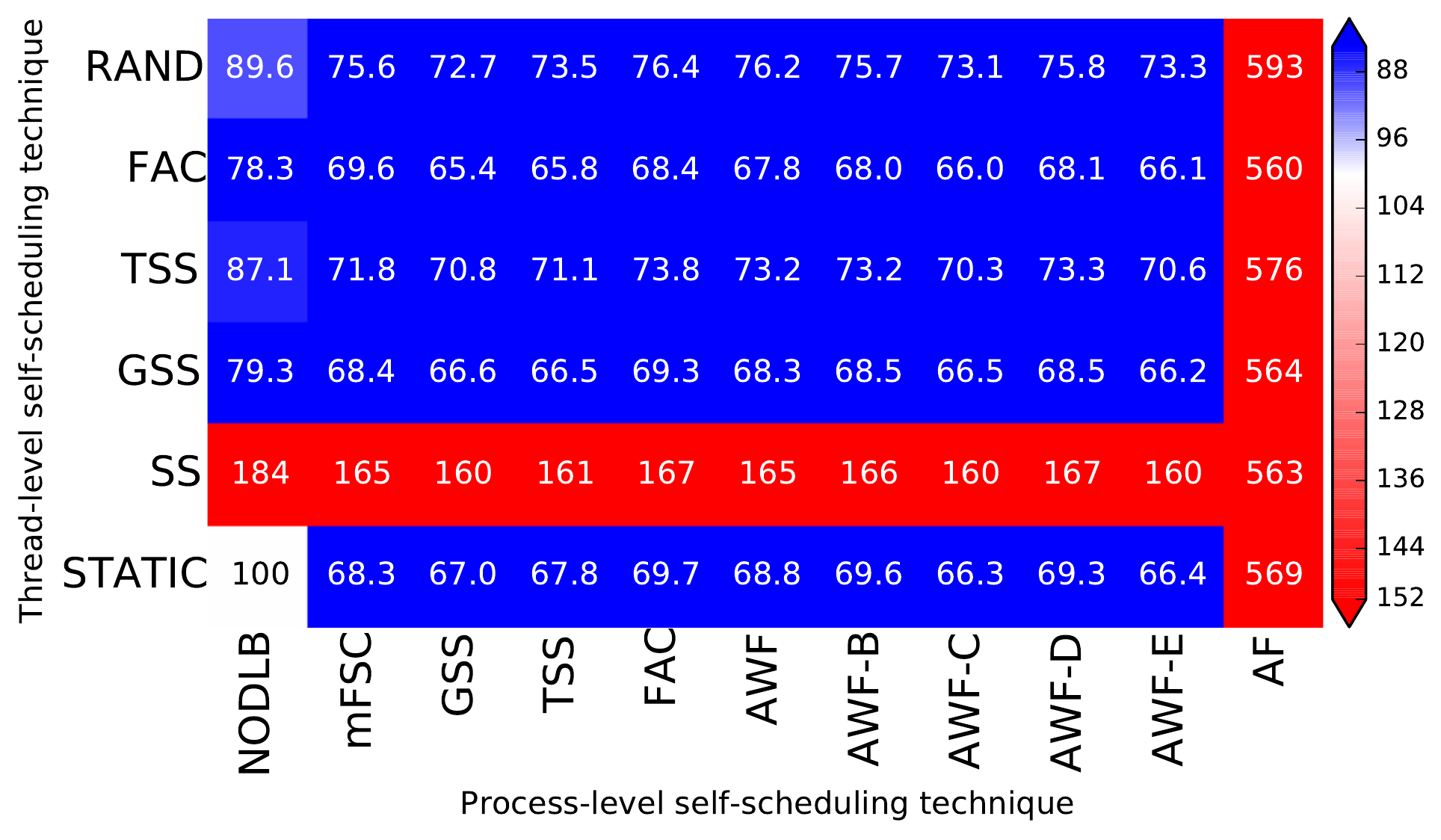}%
		\label{subfig:evrard_1000_heatmap}%
	} \\
	\caption{
		Impact of the two-level dynamic load balancing on the performance of the three scientific applications. \ali{Percent improvement} corresponds to the average of $20$ repetitions or \mbox{time-steps} with two-level load balancing normalized with respect to NODLB\_STATIC. White, red and blue correspond to baseline, degraded and improved performance, respectively.
	}
	\label{fig:heat_map}
\end{figure}
\clearpage
\begin{figure}[p!]
	\centering
	\subfloat[Evrard collapse, $1M$ particles, time-step 1700-1720, $12$ processes, $10$ threads (miniHPC)]{%
		\includegraphics[clip, trim=0cm 0cm 0cm 0cm,scale=0.43]{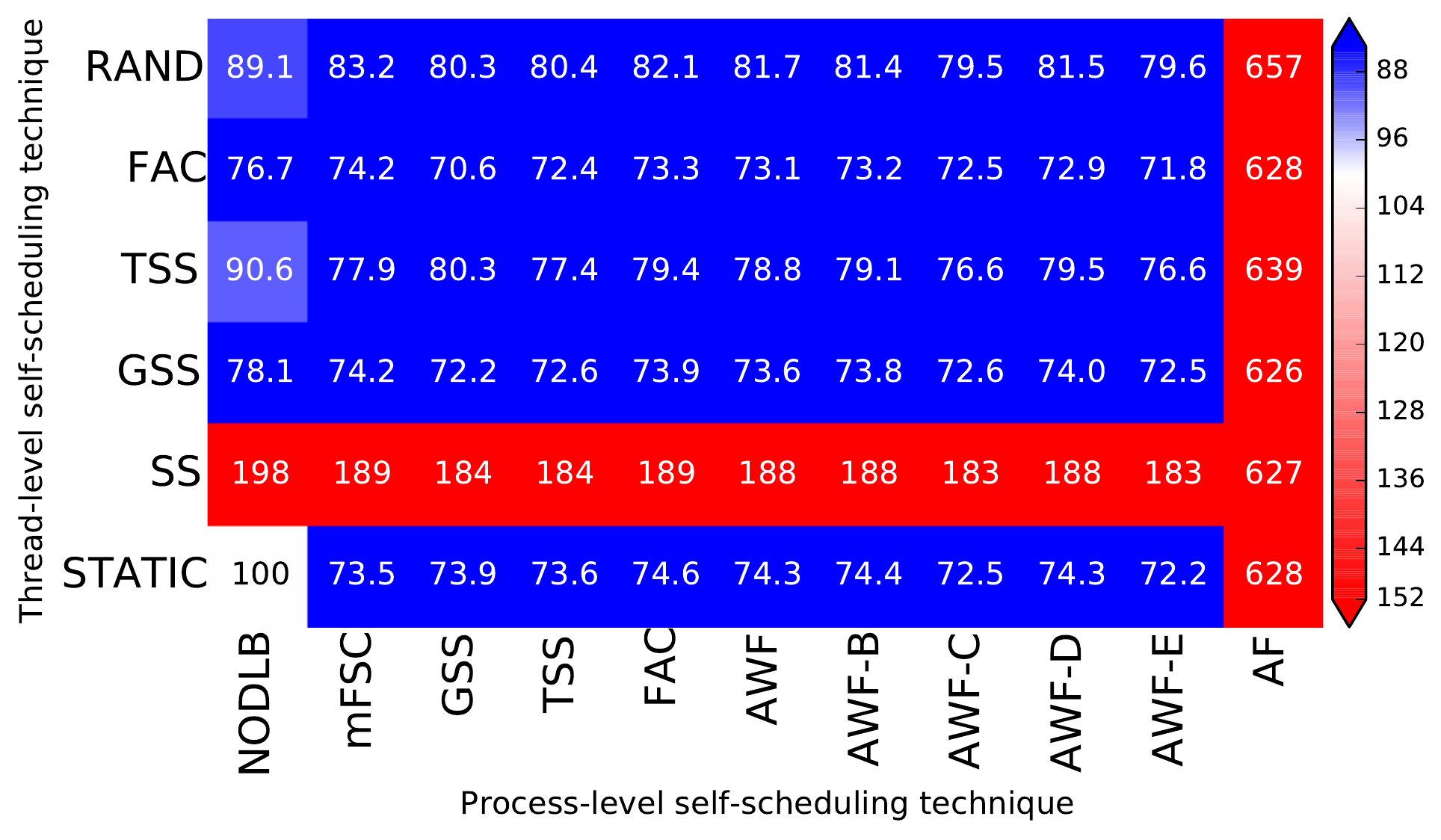}
		\label{subfig:evrard_1700_heatmap}%
	} \hspace{0cm}
	\subfloat[Evrard collapse, $1M$ particles, time-step 2000-2020, $12$ processes, $10$ threads (miniHPC)]{%
		\includegraphics[clip, trim=0cm 0cm 0cm 0cm, scale=0.43]{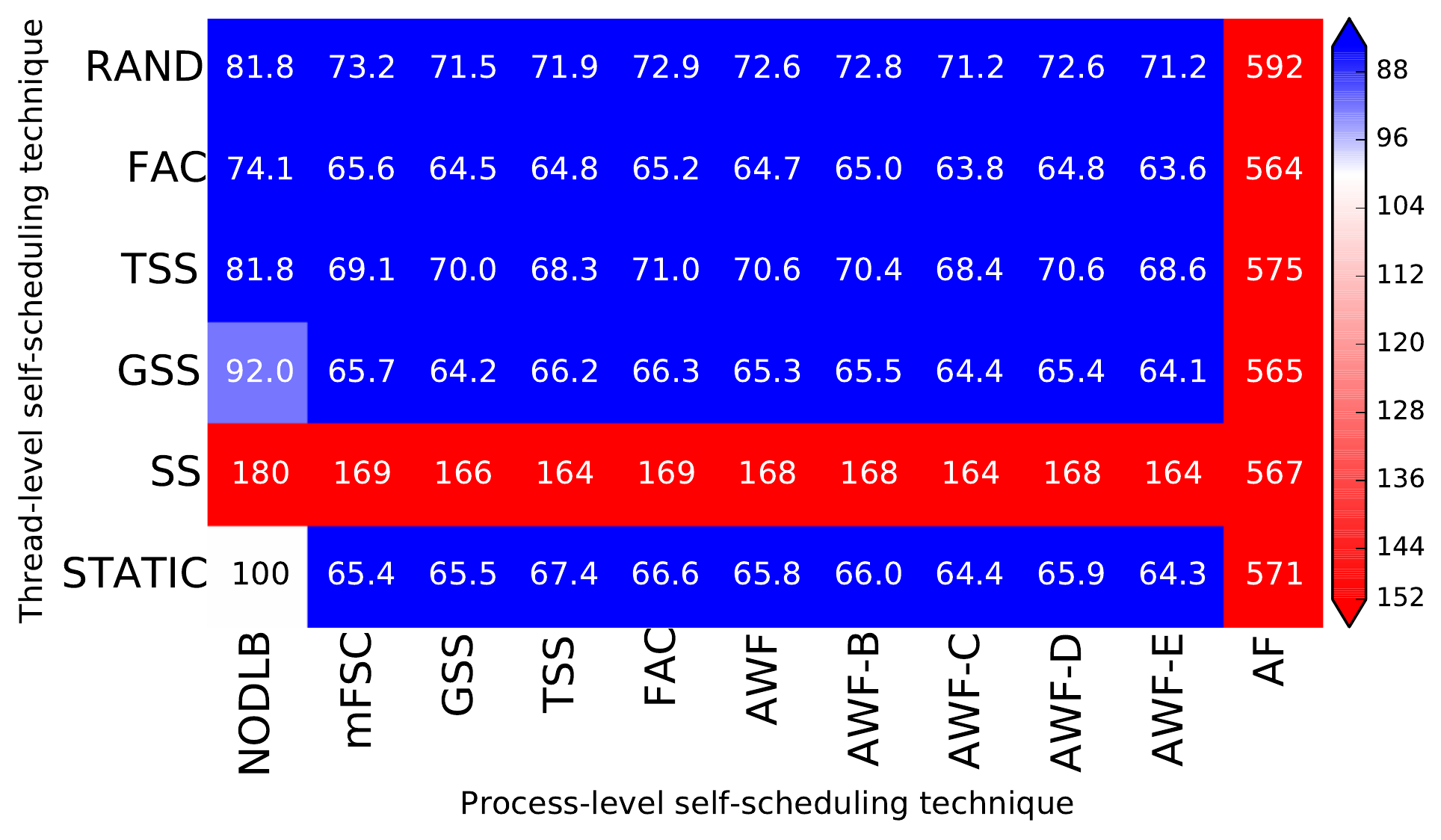}%
		\label{subfig:evrard_2000_heatmap}%
	} \\
	\subfloat[Evrard collapse, $1M$ particles, time-step 2300-2320, $12$ processes, $10$ threads (miniHPC)]{%
		\includegraphics[clip, trim=0cm 0cm 0cm 0cm,scale=0.43]{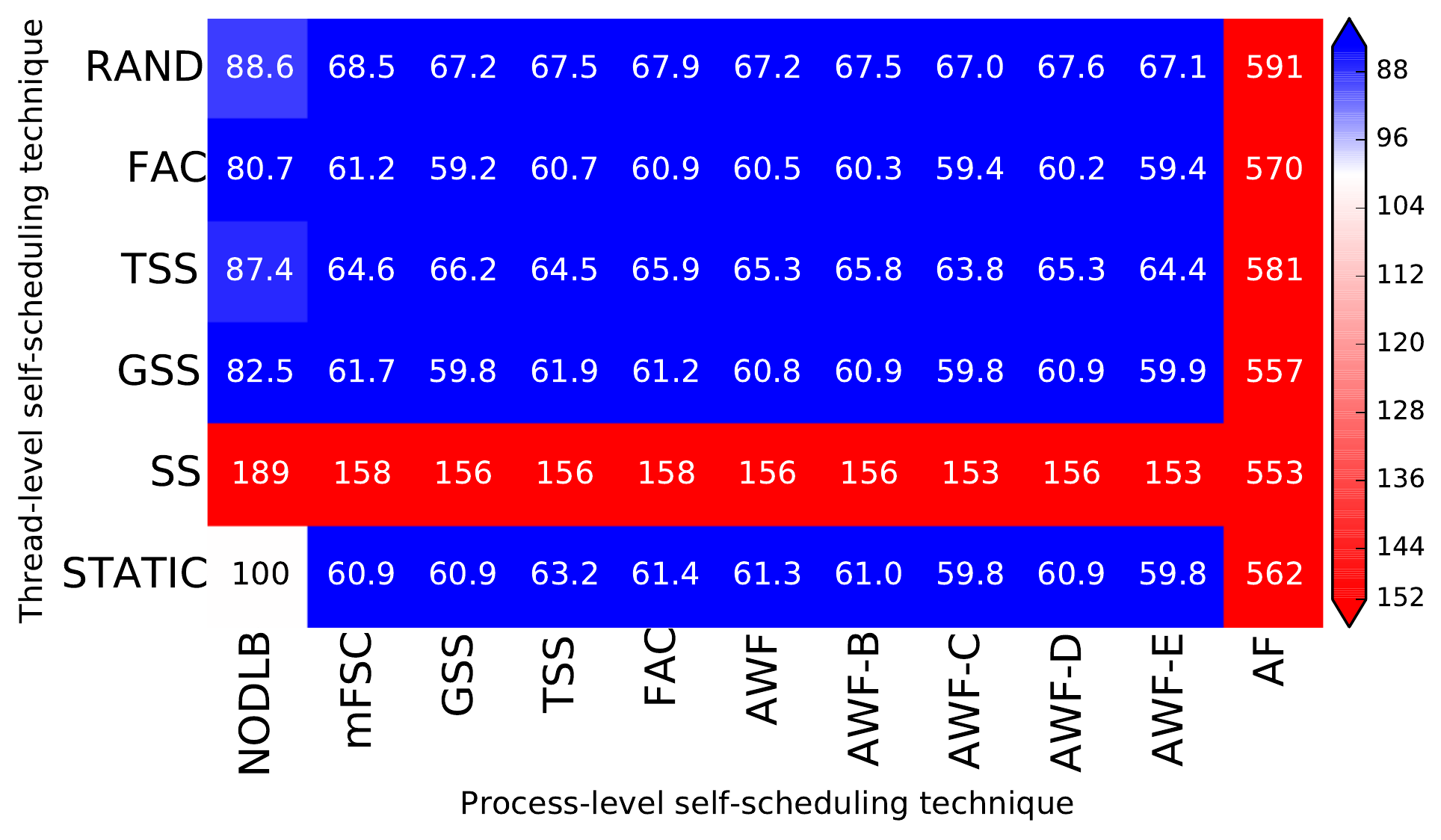}
		\label{subfig:evrard_2300_heatmap}%
	} \hspace{0cm}
	\subfloat[Evrard collapse, $1M$ particles, time-step 2500-2520, $12$ processes, $10$ threads (miniHPC)]{%
		\includegraphics[clip, trim=0cm 0cm 0cm 0cm, scale=0.43]{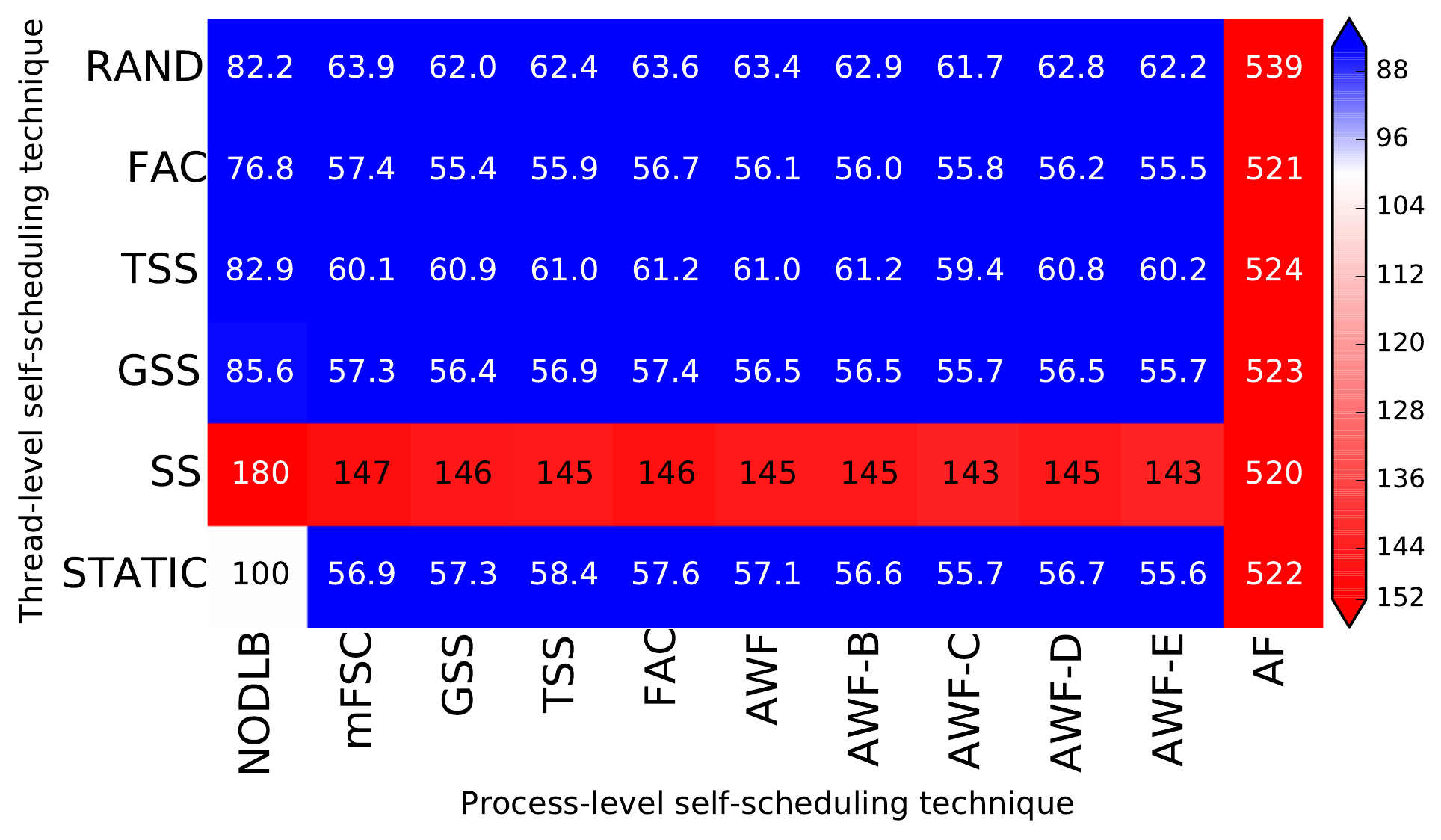}%
		\label{subfig:evrard_2500_heatmap}%
	} \\
	\subfloat[Evrard collapse, $1M$ particles, time-step 2800-2820, $12$ processes, $10$ threads (miniHPC)]{%
		\includegraphics[clip, trim=0cm 0cm 0cm 0cm, scale=0.43]{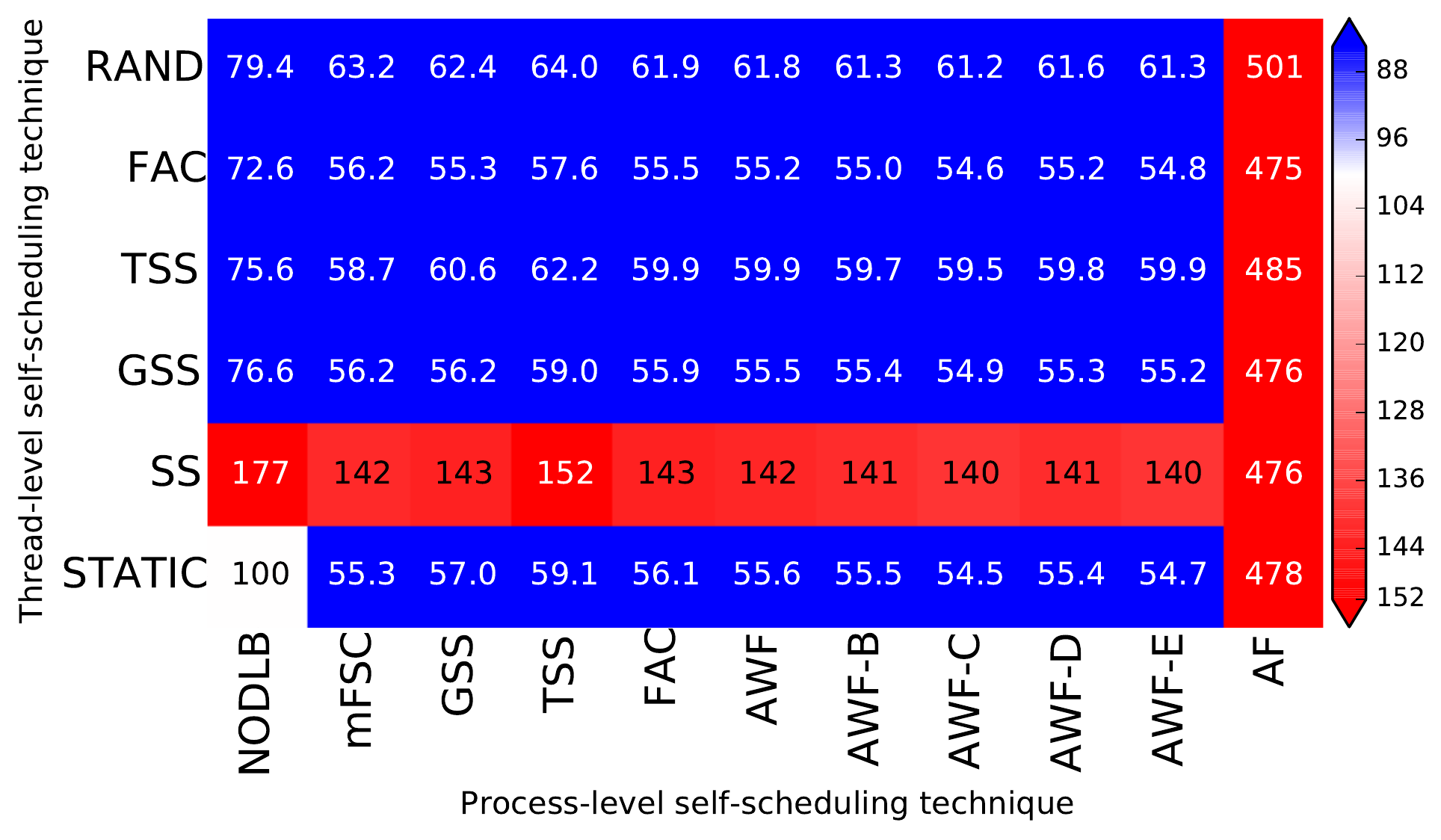}%
		\label{subfig:evrard_2800_heatmap}%
	} \hspace{0cm}
	\subfloat[Evrard collapse (gravity) $10M$ particles, time-steps 1-20, $100$ processes, $12$ threads (Piz Daint)]{%
		\includegraphics[clip, trim=0cm 0cm 0cm 0cm, scale=0.43]{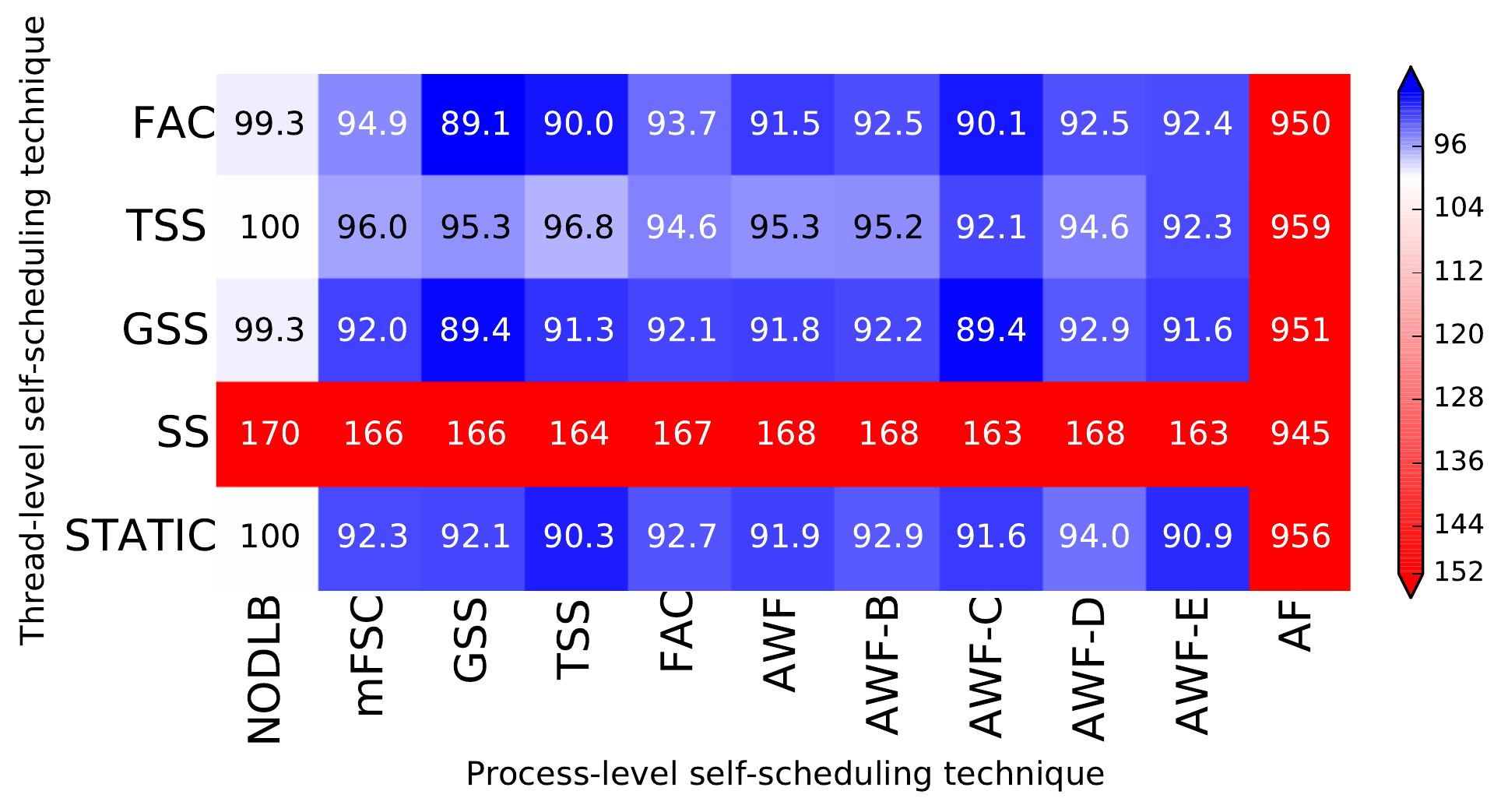}%
		\label{subfig:evrard_1_heat_map}%
	} \\
	\caption{
		Impact of the two-level dynamic load balancing on the performance of the Evrard collapse \mbox{test-case}, $1M$ and $10M$ particles, with \sphynx{}. \ali{Percent improvement} corresponds to the average of $20$ repetitions or \mbox{time-steps} with \mbox{two-level} load balancing normalized with respect to NODLB\_STATIC. White, red and blue correspond to baseline, degraded and improved performance, respectively.
	}
	\label{fig:heat_map2}
\end{figure}
\clearpage

\begin{figure}[p!]
	\centering
	\subfloat[Mandelbrot, $40$ processes, $10$ threads (miniHPC)]{%
		\includegraphics[clip, trim=0cm 0cm 0cm 0cm,scale=0.22]{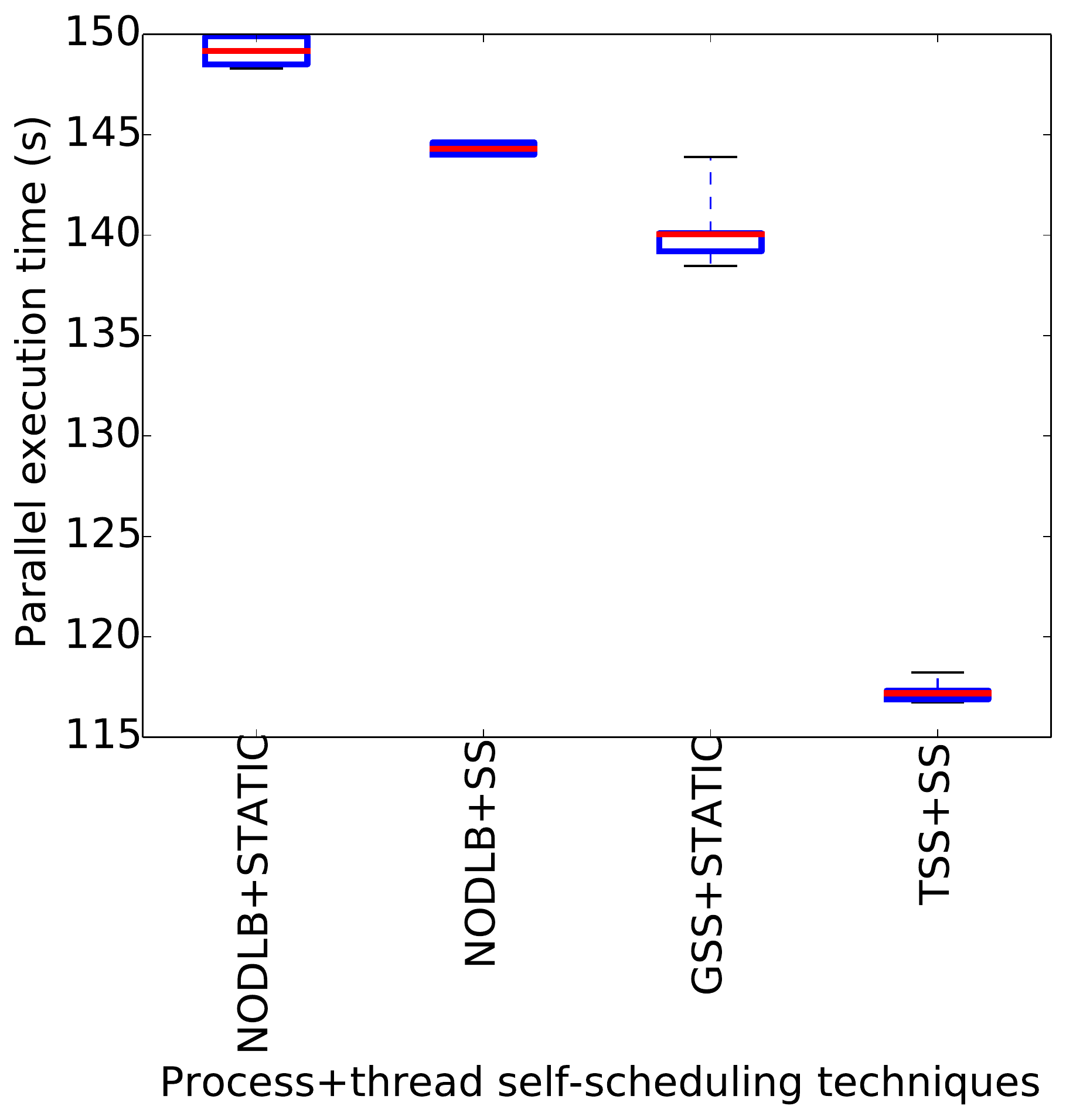}
		\label{subfig:Mandelbrot_per_level}%
	} \hspace{0cm} 
	\subfloat[PSIA, $40$ processes, $10$ threads (miniHPC)]{%
		\includegraphics[clip, trim=0cm 0cm 0cm 0cm, scale=0.22]{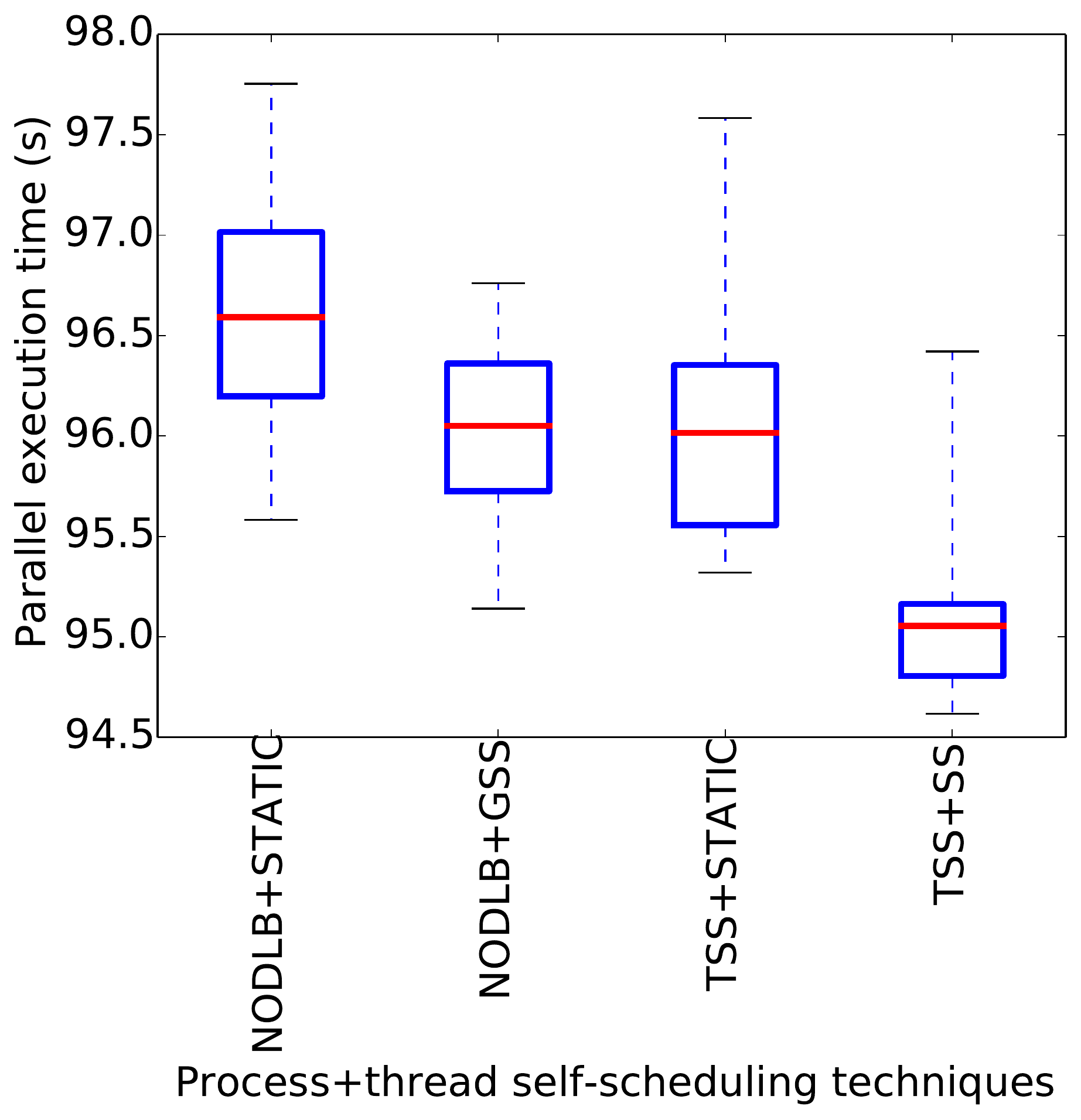}%
		\label{subfig:PSIA_per_level}%
	} \hspace{0cm}
	\subfloat[Stellar collision, $10M$ particles, time-step 6900-6920, $40$ processes, $10$ threads (miniHPC)]{%
		\includegraphics[clip, trim=0cm 0cm 0cm 0cm,scale=0.22]{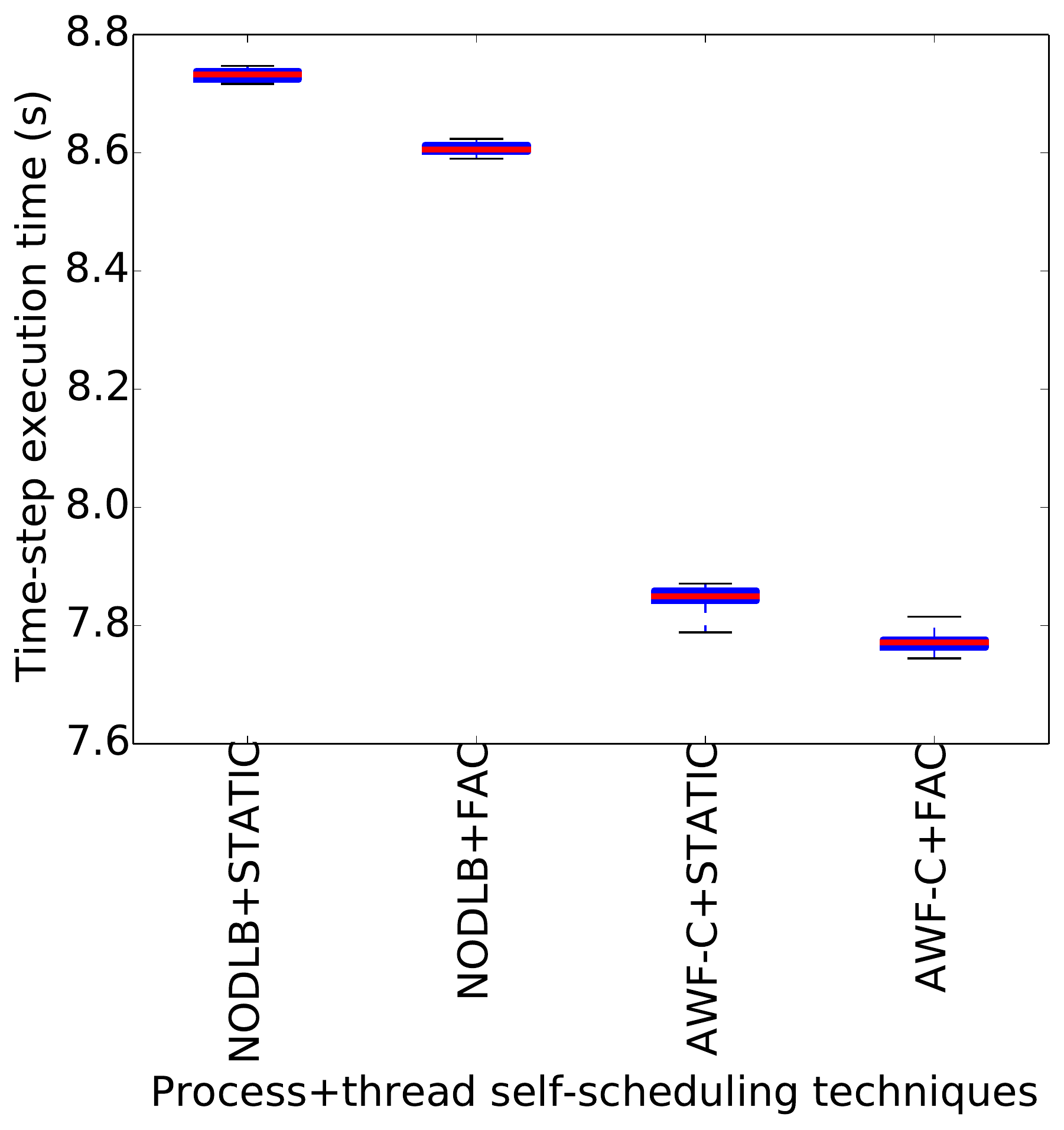}
		\label{subfig:collision_per_level}%
	} \hspace{0cm} 
	\subfloat[Evrard collapse, $1M$ particles, time-step 100-120, $12$ processes, $10$ threads (miniHPC)]{%
		\includegraphics[clip, trim=0cm 0cm 0cm 0cm, scale=0.22]{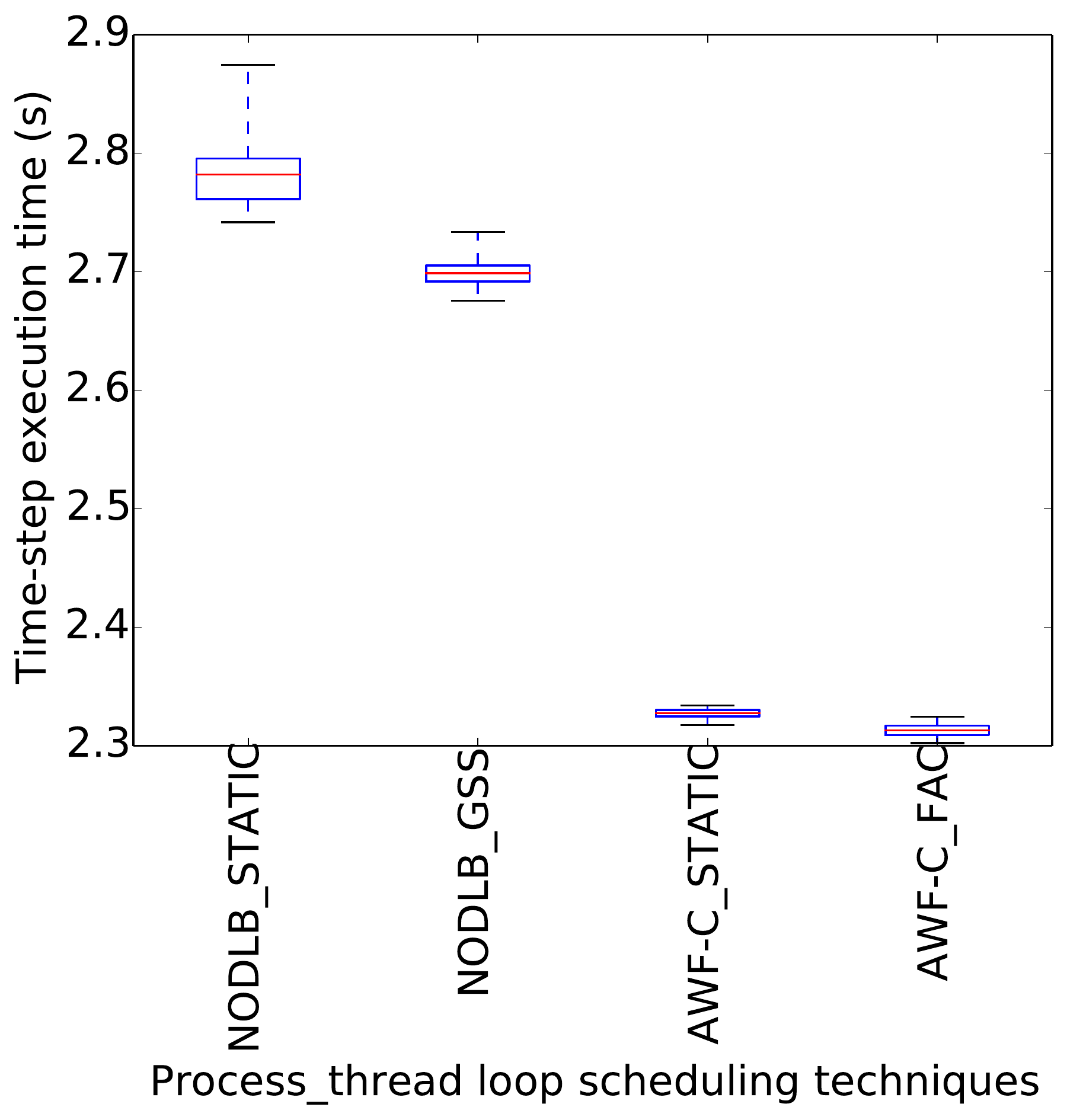}%
		\label{subfig:evrard_100_per_level}%
	} \\
	\subfloat[Evrard collapse, $1M$ particles, time-step 500-520, $12$ processes, $10$ threads (miniHPC)]{%
		\includegraphics[clip, trim=0cm 0cm 0cm 0cm, scale=0.22]{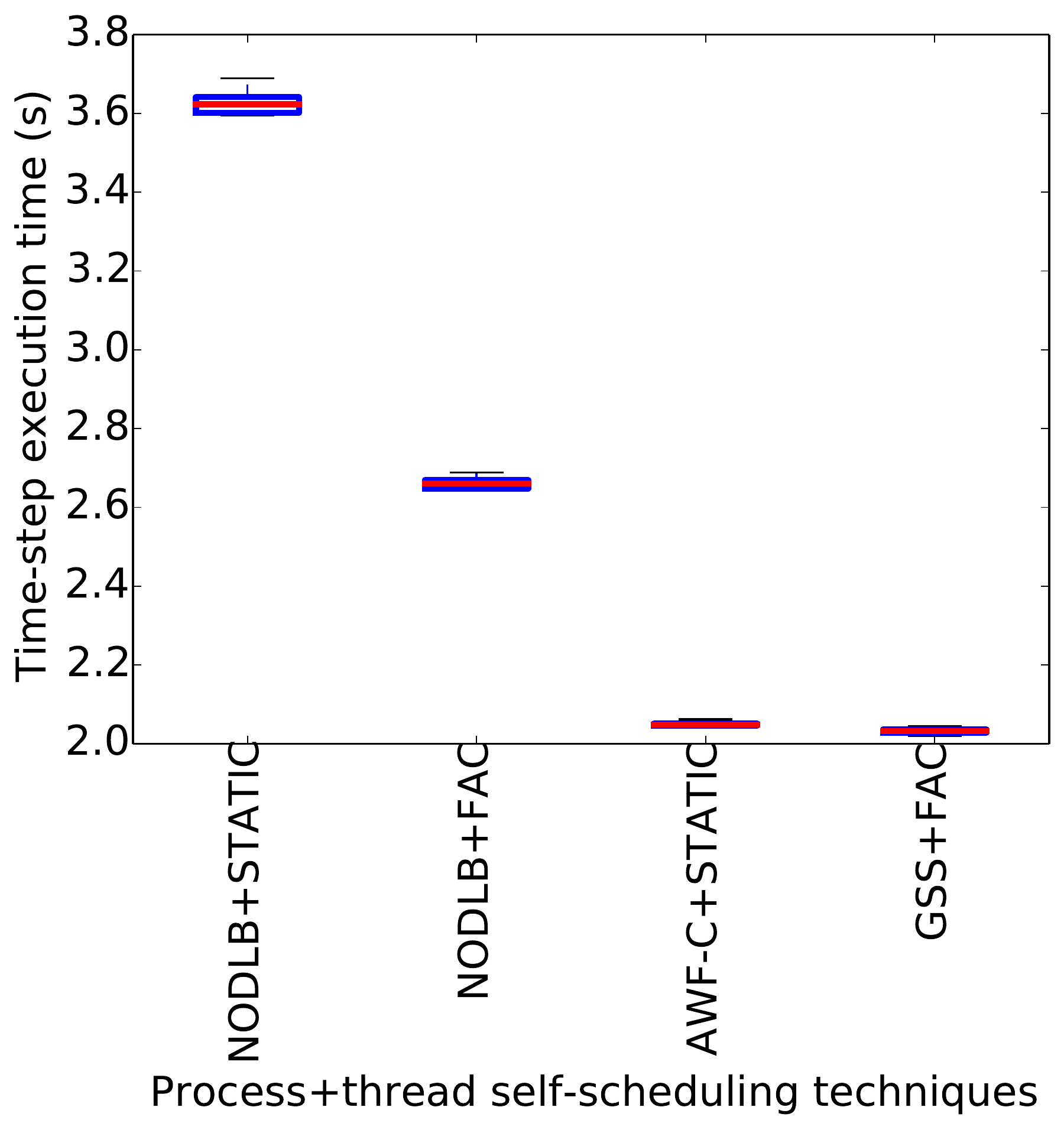}%
		\label{subfig:evrard_500_per_level}%
	} \hspace{0cm} 
	\subfloat[Evrard collapse, $1M$ particles, time-step 1000-1020, $12$ processes, $10$ threads (miniHPC)]{%
		\includegraphics[clip, trim=0cm 0cm 0cm 0cm, scale=0.22]{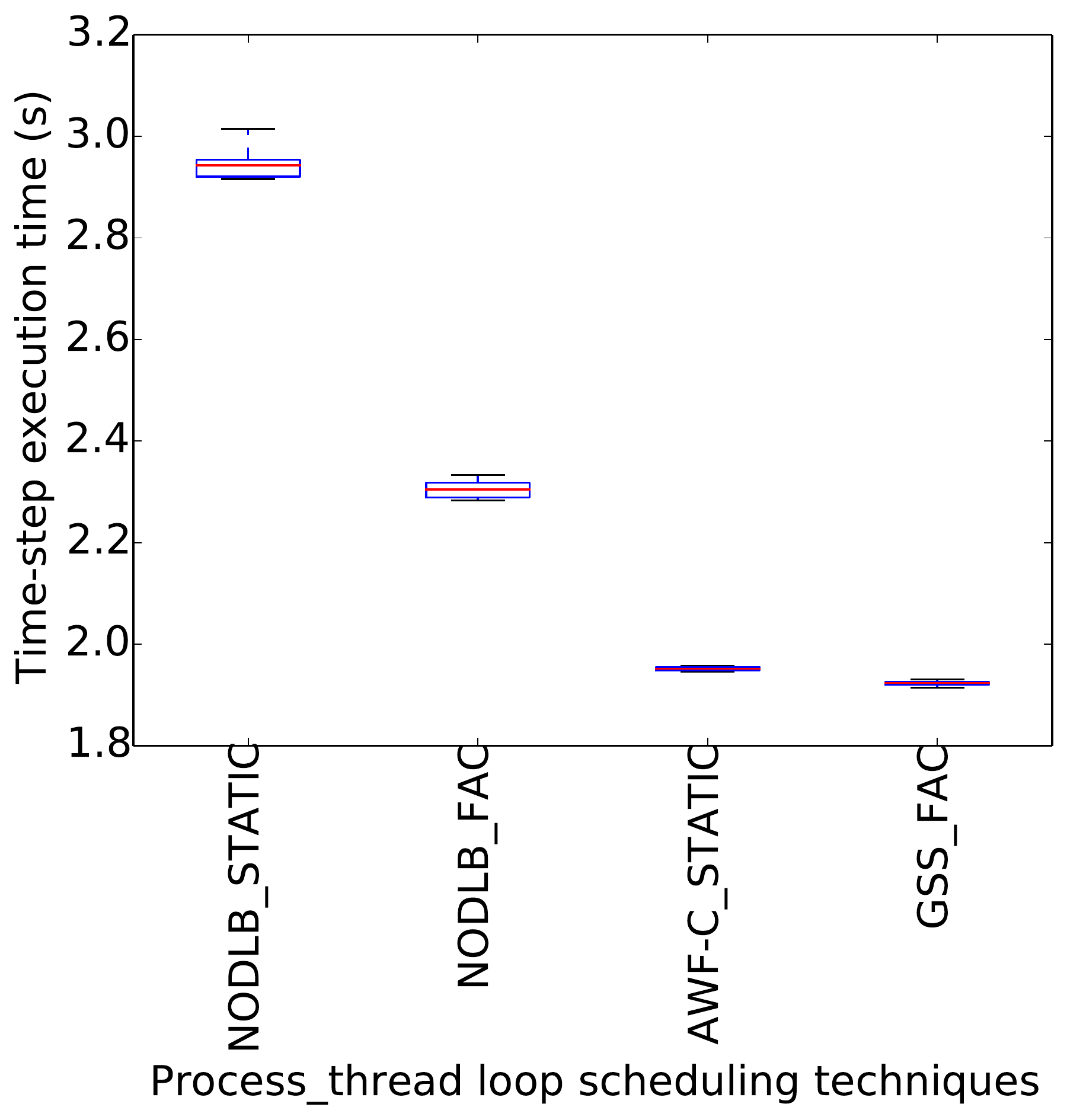}%
		\label{subfig:evrard_1000_per_level}%
	} \hspace{0cm} 
	\subfloat[Evrard collapse, $1M$ particles, time-step 1700-1720, $12$ processes, $10$ threads (miniHPC)]{%
		\includegraphics[clip, trim=0cm 0cm 0cm 0cm, scale=0.22]{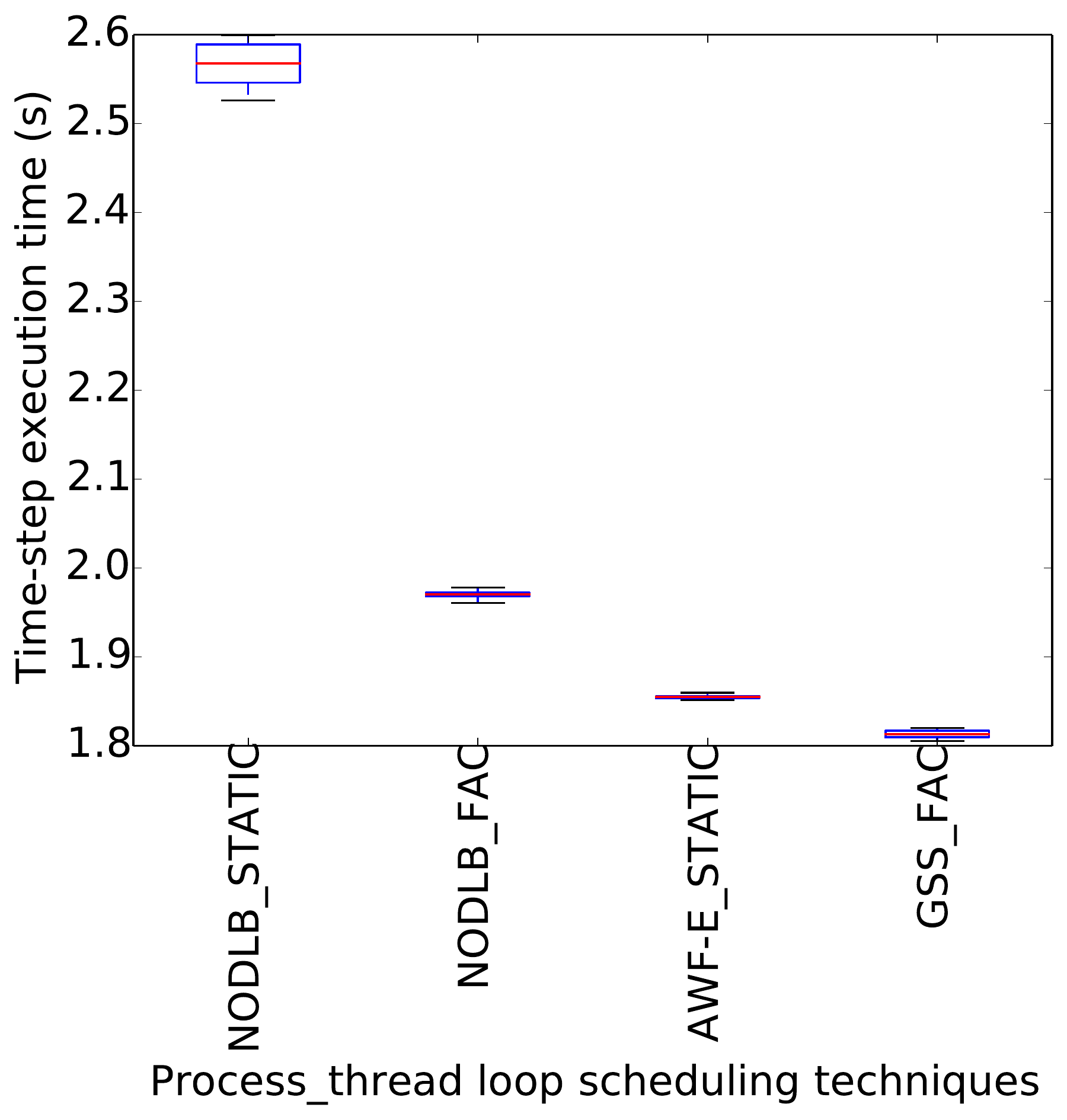}%
		\label{subfig:evrard_1700_per_level}%
	} \hspace{0cm} 
	\subfloat[Evrard collapse, $1M$ particles, time-step 2000-2020, $12$ processes, $10$ threads (miniHPC)]{%
		\includegraphics[clip, trim=0cm 0cm 0cm 0cm, scale=0.22]{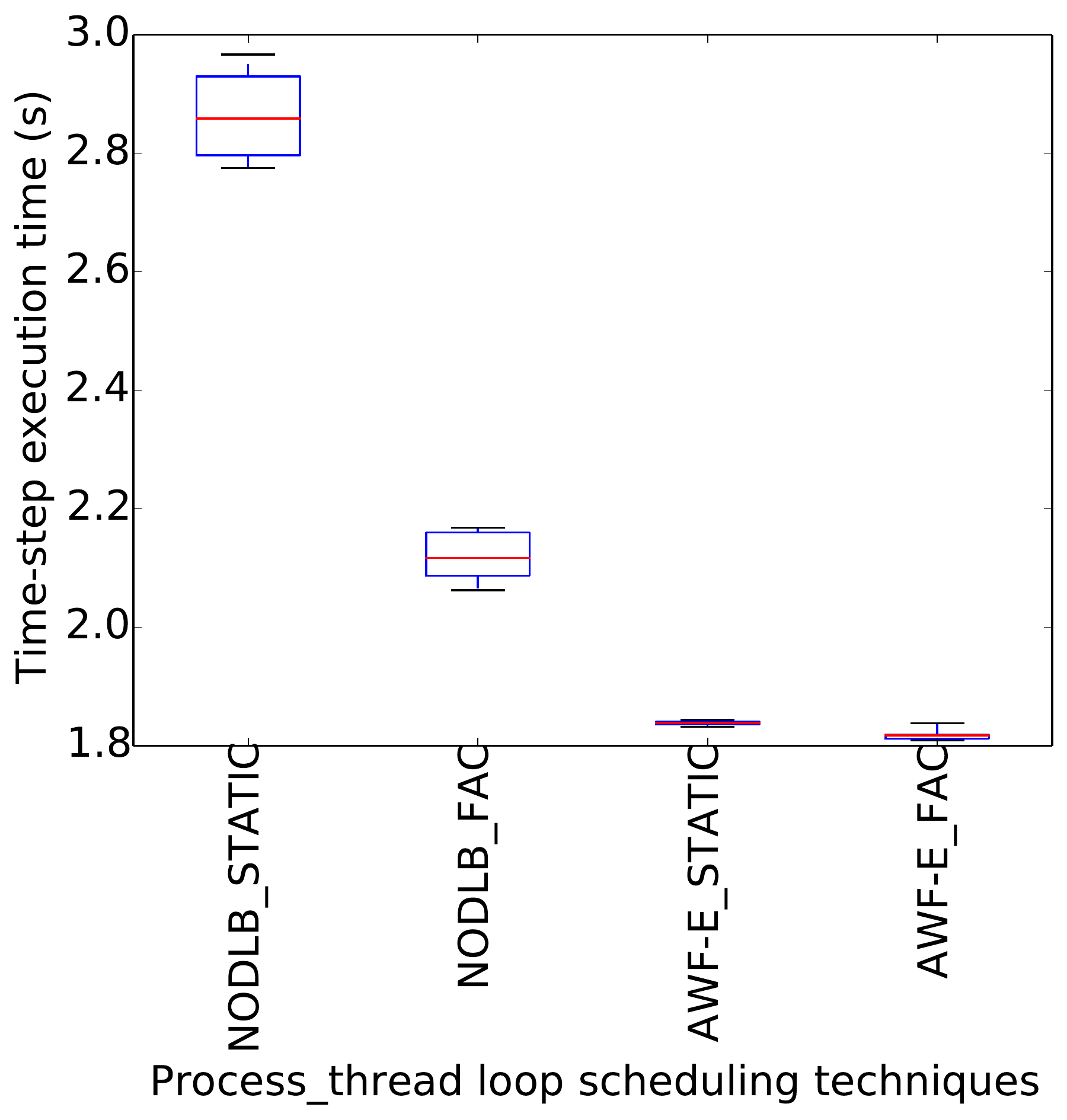}%
		\label{subfig:evrard_2000_per_level}%
	} \\
	\subfloat[Evrard collapse, $1M$ particles, time-step 2300-2320, $12$ processes, $10$ threads (miniHPC)]{%
		\includegraphics[clip, trim=0cm 0cm 0cm 0cm, scale=0.22]{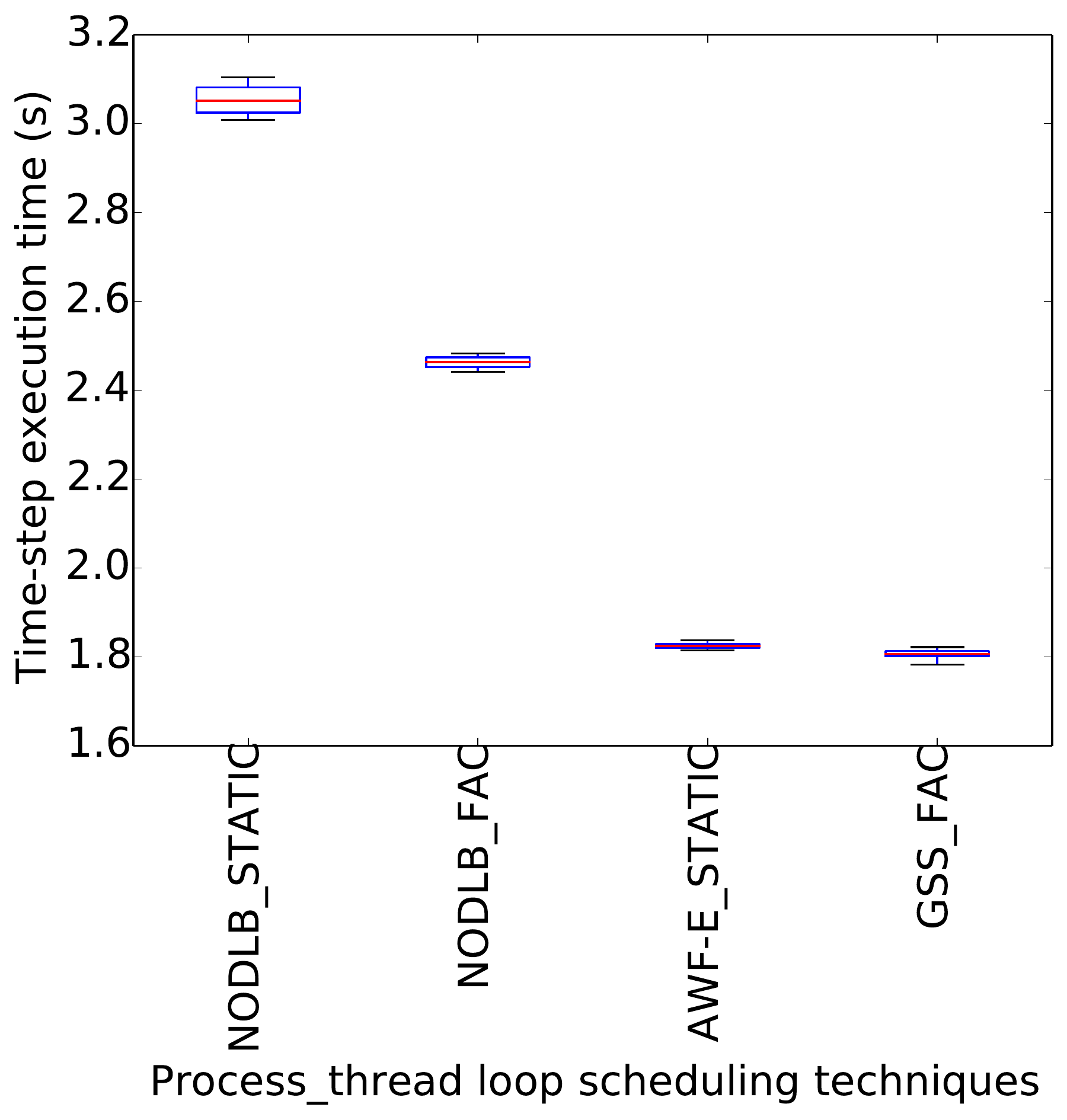}%
		\label{subfig:evrard_2300_per_level}%
	} \hspace{0cm} 
	\subfloat[Evrard collapse, $1M$ particles, time-step 2500-2520, $12$ processes, $10$ threads (miniHPC)]{%
		\includegraphics[clip, trim=0cm 0cm 0cm 0cm, scale=0.22]{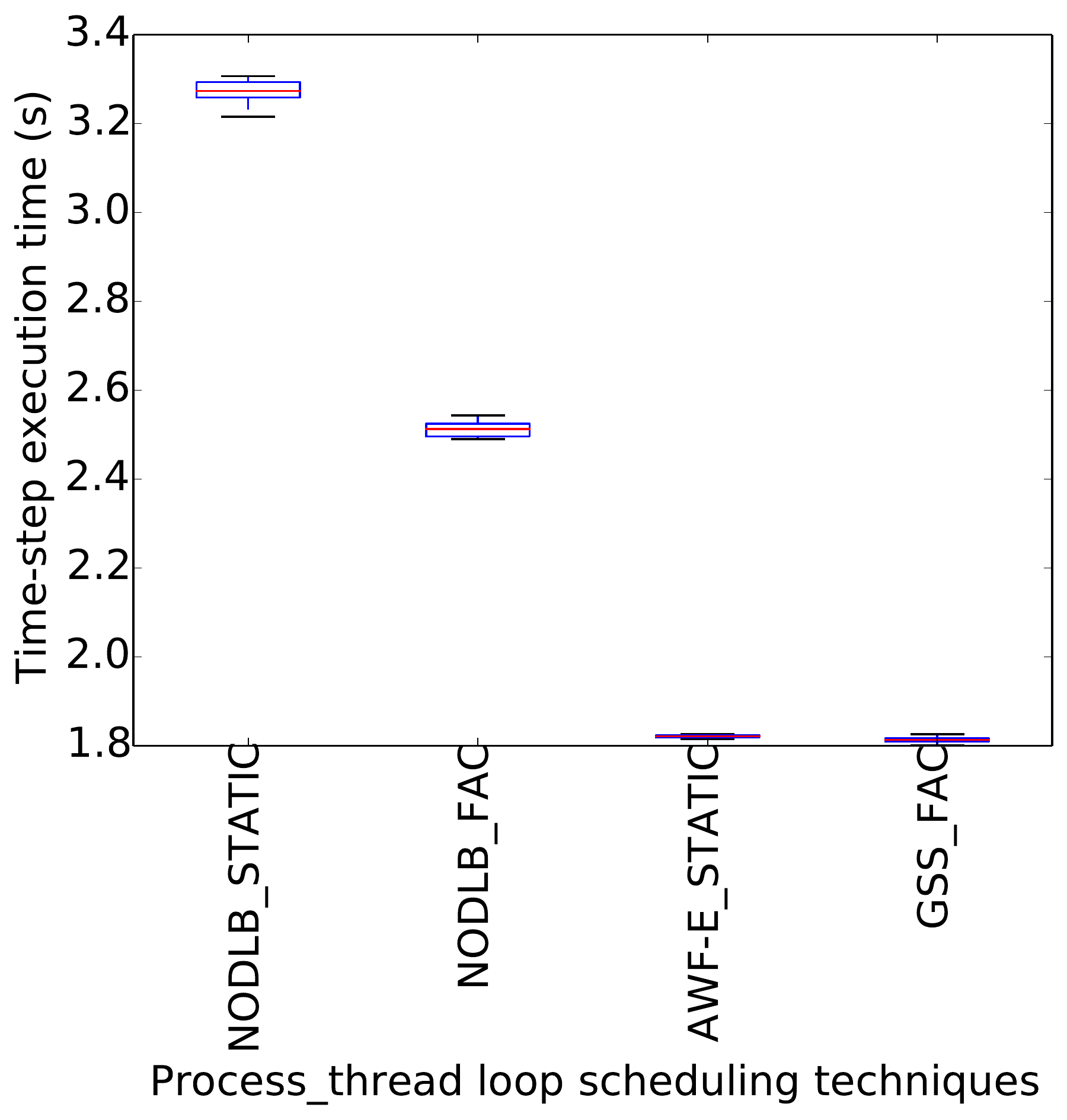}%
		\label{subfig:evrard_2500_per_level}%
	} \hspace{0cm} 
	\subfloat[Evrard collapse, $1M$ particles, time-step 2800-2820, $12$ processes, $10$ threads (miniHPC)]{%
		\includegraphics[clip, trim=0cm 0cm 0cm 0cm, scale=0.22]{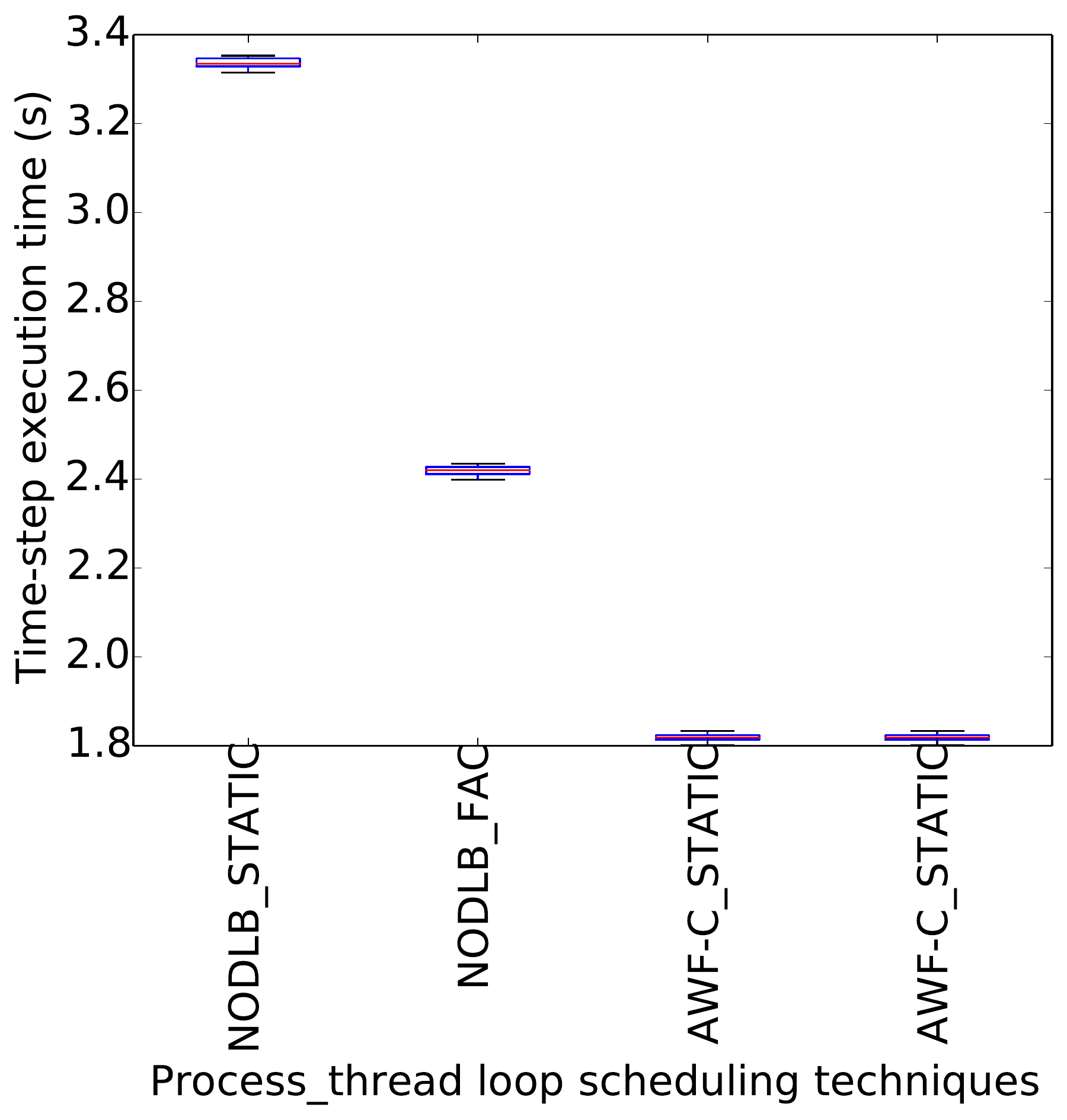}%
		\label{subfig:evrard_2800_per_level}%
	} \hspace{0cm} 
	\subfloat[Evrard collapse (gravity), $10M$ particles, time-steps 1-20, $100$ processes, $12$ threads (Piz~Daint)]{%
		\includegraphics[clip, trim=0cm 0cm 0cm 0cm, scale=0.22]{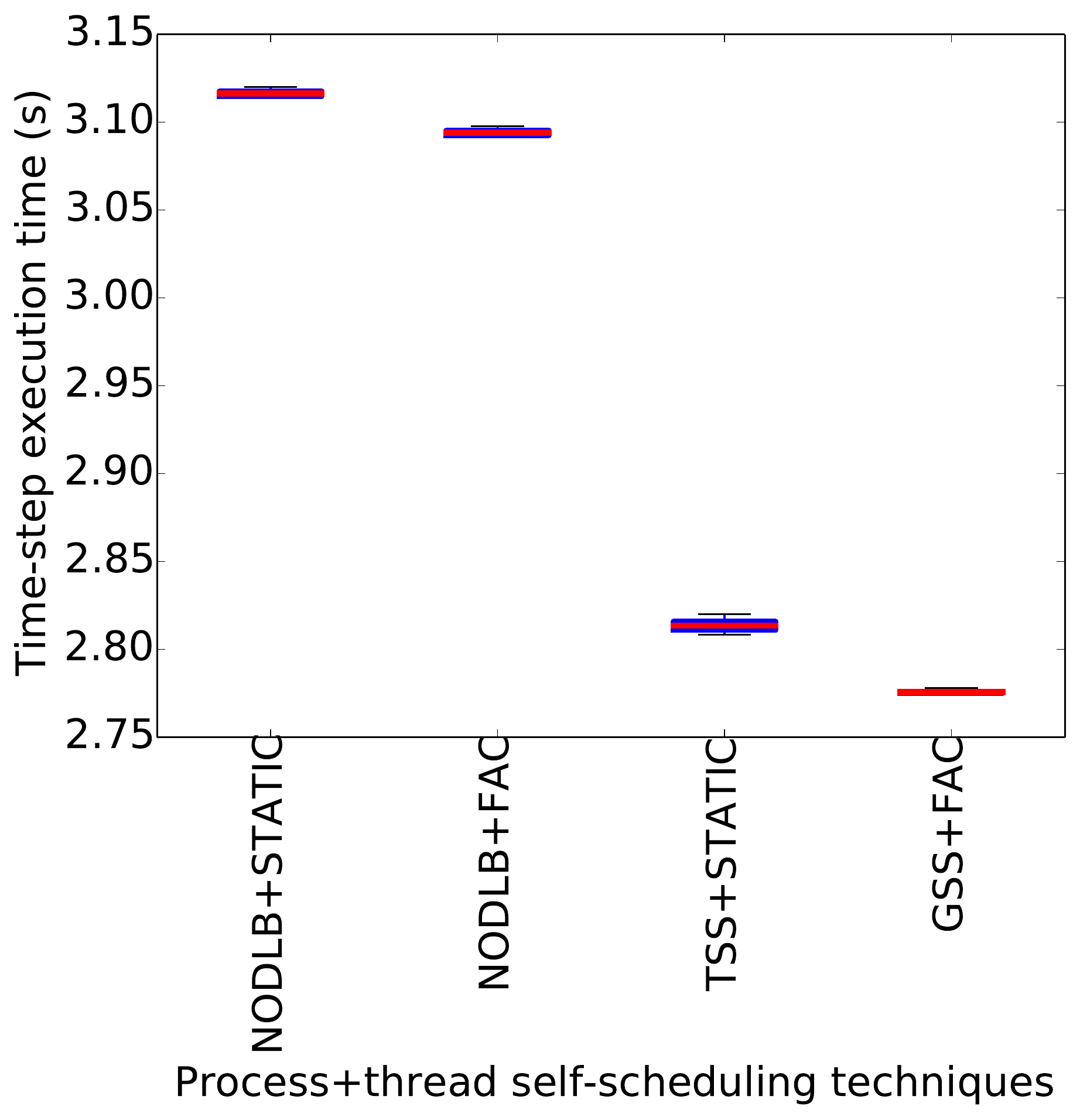}%
		\label{subfig:evrard_1_per_level}%
	}\\
	\caption{
		Impact of single and two-level dynamic load balancing on the execution time of the three scientific applications. Each plot shows in the following order: the execution time with the baseline (NODLB\_STATIC), best DLS technique at \tl{}, best DLS technique at \pl{}, and best two-level combination. The red line represents the average performance over $20$ repetitions or~\mbox{time-steps}, the boxes define the first and third quartiles, and the whiskers are maximum and minimum values.
	}
	\label{fig:per_level_performance}
\end{figure}
\clearpage

\subsection{Discussion}
Execution traces in~\figurename{~\ref{fig:load_imbalance}} show different profiles of \mbox{two-level} load imbalance.
The Mandelbrot execution trace in~\figurename{~\ref{subfig:Mandelbrot_imbalance}} shows a severe case of \mbox{two-level} load imbalance, where there is high variability in processes finishing times, and in threads finishing times within a single process, as demonstrated in~\figurename{~\ref{fig:two-level-hl}}.
\mbox{single-level} load balancing (\tl{} or \pl{}) in this case achieves limited performance improvement as predicted in~\figurename{~\ref{fig:two-level}} and confirmed by experimental results in~\figurename{~\ref{subfig:Mandelbrot_per_level}}, which showed the significant improvement of performance with \mbox{two-level} dynamic load balancing versus its slight improvement with \mbox{single-level} load balancing.
While Mandelbrot represents an extreme load imbalance, PSIA represents the other extreme, where processes and threads within a process are slightly imbalanced as shown in~\figurename{~\ref{subfig:PSIA_imbalance}}.
Therefore, the maximum achievable performance improvement in PSIA is much smaller ($1.5\%$) than that in Mandelbrot~($21\%$).

\sphynx{} execution in stellar collision execution suffers from high load imbalance at the \pl{} and low load imbalance at the \tl{} as shown in~\figurename{~\ref{subfig:collision_imbalance}}.
This is reflected by the slight effect of \tl{} load balancing and the significant impact of \pl{} load balancing in improving its performance, as shown in~\figurename{~\ref{subfig:collision_per_level}}.
For the Evrard collapse, $1M$ particles, however, one can observe a severe load imbalance at the \pl{}. 
This load imbalance at the \pl{} is in fact caused by two threads lagging the execution of the last process behind all other processes as can be seen in~\figurename{~\ref{subfig:evrard_500_imbalance}}.
In this case, \tl{} load balancing not only improves the load balancing at the \tl{} but also at the \pl{} as the last process will finish early, making it closer to the other processes finishing times.

\begin{figure}[t]
	\centering
	\includegraphics[clip, trim=0cm 0cm 0cm 0cm, scale=0.56]{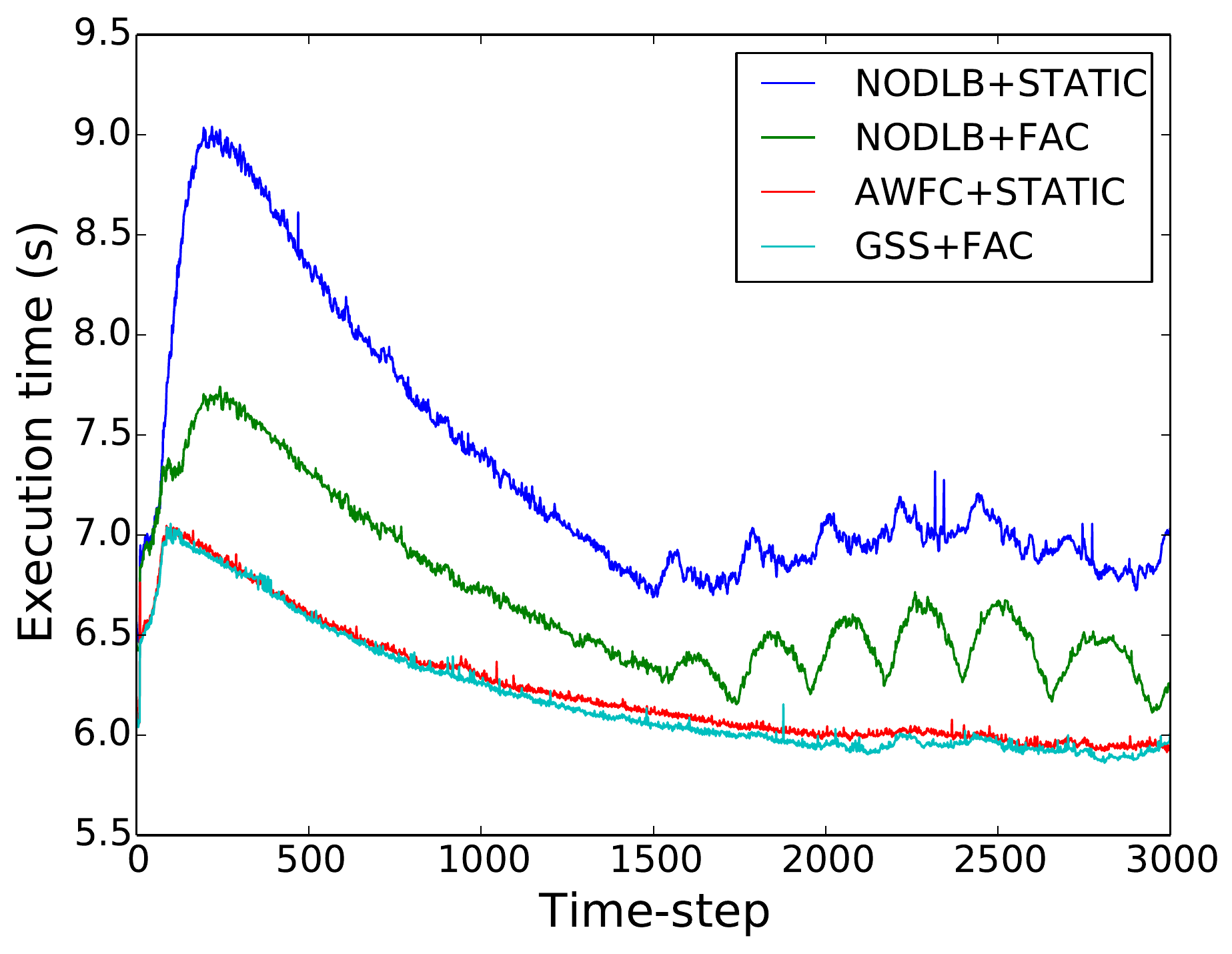}%
	\caption{Time-step execution time of Evrard collapse test-case, $1M$ particles, in \sphynx{} with 12 processes and 10 threads per process on miniHPC. The \mbox{two-level} dynamic load balancing improved the performance of \sphynx{} throughout the full simulation, achieving $15\%$ overall improvement in execution time.}
	\label{fig:full_simulation}
\end{figure}
\clearpage

Alternatively, \pl{} load balancing will also distribute the high workload of the last two threads among all the processes, therefore, dissolving the \tl{} load imbalance among processes.
This is confirmed by the significant performance improvement by both \tl{} alone and \pl{} alone in~\figurename{~\ref{subfig:evrard_500_per_level}}.
This represents an interesting case where load balancing from the \tl{} propagates to the \pl{} and vice versa.

Another important observation from \figurename{~\ref{fig:per_level_performance}} is that the best combination of DLS techniques at the \tl{} and \pl{} (fourth column), is not always the combination of the best technique at the \tl{} (second column) and the best technique at the \pl{} (third column).
These show the interplay between the \tl{} and the \pl{} load balancing, as load balancing at one level changes the load imbalance at the other level. 
As illustrated in \figurename{~\ref{fig:two-level} the load balancing techniques \ali{are needed} at both levels \ali{in the right combination}, to achieve \ali{the best} balanced load execution.

\section{Conclusion and Future Work} 
\label{sec:conc}
%

In this work, load imbalance at the \tl{} and \pl{} of \aliA{three scientific applications,} 
\ac{has been} analyzed.
Load imbalance degrades performance, adversely affects scalability, and becomes more significant as the number of processes increases.
\aliA{Dynamic load balancing via \mbox{self-scheduling} has been used in this work to address the \mbox{two-level} load imbalance and improve scientific applications performance.}
\aliA{\elap{} has been used to employ DLS at the \tl{} and the \dlbTool{} at the \pl{}.}
\ac{In addition, the} \dlbTool{} \ac{has been} extended with \ali{the} AWF technique that \ali{is specifically designed for} time-stepping scientific applications, such as \sphynx{}. 
The proposed \mbox{two-level} load balancing \ali{approach} using \ali{the} \dlbTool{} and \ali{the} \elap{} is \emph{generic} and can be applied to any MPI+OpenMP application.
\aliA{Based on the nature of the load imbalance at the \tl{} and \pl{}, certain performance improvements can be achieved with \mbox{single-level} load balance either at the \tl{} or the \pl{}.}
\aliA{However, the best} application performance can only be achieved by addressing load imbalance \emph{jointly} at the \mbox{two-levels}.
Also, the DLS techniques at \tl{} and \pl{} \ali{influence each other}, \ali{and this influence should not be ignored.} 
\aliA{In certain cases, load balancing at \tl{} propagates to the \pl{} and vice versa.}
\aliA{In addition,} the best performing \ali{two-level} combination is not always the \ac{combination of the two} best performing DLS techniques at \ali{a single level alone}.
\aliA{This highlights the interplay between \tl{} and \pl{} load balancing, as load balancing at one level changes the load imbalance at the other level.}

\aliA{The extension of the \dlbTool{} to work with distributed data as well as replicated data is planned in the future.}
\aliA{In addition, implementing \mbox{self-scheduling} techniques using decentralized control approach to improve their scalability is also envisioned in the future.}
Also, an intelligent selection of the best combination of \tl{} and \pl{} scheduling techniques using simulation or machine learning is planned as future work.

\section*{Acknowledgment}
This work has been \ali{partially} supported by the Swiss Platform for Advanced Scientific Computing (PASC) project SPH-EXA: Optimizing Smooth Particle Hydrodynamics for Exascale Computing and by the Swiss National Science Foundation in the context of the ``Multi-level Scheduling in Large Scale High Performance Computers'' (MLS) grant, number 169123.

\bibliographystyle{ieeetr}
\bibliography{citedatabase}
\end{document}